\documentclass[a4paper,10pt]{article}

\pdfoutput=1
\usepackage{jheppub}
\usepackage{array}
\usepackage{multirow}
\usepackage{amsmath}
\usepackage{makecell}

\usepackage{cancel}
\usepackage{amsmath,amssymb} 
\usepackage{graphicx} 
\usepackage[dvipsnames,x11names]{xcolor} 

\usepackage{hyperref}
\usepackage{cleveref}
\usepackage{float}
\usepackage{subcaption}

\makeatletter
\newcommand{\footnoteref}[1]{\protected@xdef\@thefnmark{\ref{#1}}\@footnotemark}
\makeatother

\title{\textbf{The zeroth law of black hole thermodynamics in arbitrary higher derivative theories of gravity}}

\author[a]{Sayantani Bhattacharyya}
\affiliation[a]{School of Physical Sciences, National Institute of Science Education and Research,\\
An OCC of Homi Bhabha National Institute, Jatni 752050, Odisha, India}

\author[b]{, Parthajit Biswas}
\affiliation[b]{Department of Physics, Indian Institute of Technology Kanpur, Kalyanpur, Kanpur 208016, India}

\author[a]{, Anirban Dinda}
 
\author[b]{, Nilay Kundu}

\emailAdd{sayanta@niser.ac.in, parthajitbiswas8@gmail.com, anirban.dinda@niser.ac.in, nilayhep@iitk.ac.in}

\abstract{We consider diffeomorphism invariant theories of gravity with arbitrary higher derivative terms in the Lagrangian as corrections to the leading two derivative theory of Einstein's general relativity. We construct a proof of the zeroth law of black hole thermodynamics in such theories. We assume that a stationary black hole solution in an arbitrary higher derivative theory can be obtained by starting with the corresponding stationary solution in general relativity and correcting it order by order in a perturbative expansion in the coupling constants of the higher derivative Lagrangian. We prove that surface gravity remains constant on its horizon when computed for such stationary black holes, which is the zeroth law. We argue that the constancy of surface gravity on the horizon is related to specific components of the equations of motion in such theories. We further use a specific boost symmetry of the near horizon space-time of the stationary black hole to constrain the off-shell structure of the equations of motion. Our proof for the zeroth law is valid up to arbitrary order in the expansion in the higher derivative couplings. 
 }
 

\begin{document}

\maketitle


\section{Introduction} \label{sec:Intro}
Black holes are fascinating gravitational objects with many properties that are very surprising and peculiar. However, black holes share one common property with ordinary matter: they also behave as thermodynamic objects. In Einstein's theory of two derivative classical gravity, also known as general relativity, they are solutions to the equations of motion, and one can associate geometric notions with corresponding thermodynamic properties, such as the area of the event horizon as its entropy, the surface gravity as the temperature, etc. It has long been understood that the laws of black hole mechanics can be viewed as laws of thermodynamics \cite{Hawking:1971tu, Bardeen:1973gs, Bekenstein:1973ur}. We also know from \cite{Hawking:1974sw} that this similarity is not an analogy. However, indeed one can derive the temperature of a black hole related to its surface gravity in a rigorous way. 

On the other hand, we know that general relativity is not a complete theory of gravity. 
We must think of it as an effective theory valid at low energies or large length scales. In a complete theory of quantum gravity, one can take its low energy limit and would get general relativity as the leading theory. Following this procedure, one would also generate corrections to general relativity. Without detailed knowledge of the UV complete theory and the process of taking a low energy limit, we cannot be sure what corrections to be added to the leading two derivative theory. Nevertheless, on general grounds, we expect that one would get various higher derivative terms in the Lagrangian in addition to the Einstein-Hilbert piece. Different higher derivative corrections will come with different dimensionful parameters as coefficients in the Lagrangian, and this will signify the length scale, say $l_{\text{HD}}$ at which the higher derivative terms would be as important as the leading Einstein gravity piece. We denote the higher derivative couplings collectively as the dimensionless parameter $\alpha$ \footnote{The higher derivative coupling will have dimensions in general, however, with the use of appropriate powers of $l_{\text{HD}}$ we can define a dimensionless coupling $\alpha$ and also choose units by putting $l_{\text{HD}} = 1$.}. 

Once we extend the scope of gravity theories by including arbitrary higher derivative corrections in addition to the leading two derivative theory, the black holes still remain to be solutions of these new theories, and they should also retain their thermodynamic properties. Therefore although the laws of black hole mechanics were first understood in general relativity, one can not ignore the importance of understanding the validity of a similar set of laws for black hole thermodynamics in such higher derivative theories of gravity. In \cite{PhysRevD.48.R3427, Iyer:1994ys}, it was shown that a version of the first law of black hole thermodynamics could indeed be argued for an arbitrary diffeomorphism invariant theory of gravity. This construction also suggested a geometric object defined on the horizon of the black hole as the generalized definition of black hole entropy. This definition of black hole entropy is known as the Wald entropy in the literature. It says that the Noether charge associated with the Killing symmetry generator of the null horizon should be identified as the entropy of black holes in such arbitrary diffeomorphism invariant theory of gravity. Of course, this definition of entropy reduces to the area of the horizon as one considers black holes in general relativity. However, once out-of-equilibrium dynamic processes involving black holes are considered, the Wald entropy suffers from possible ambiguities known as the JKM ambiguities \cite{Jacobson:1993xs, Jacobson:1993vj, Jacobson:1995uq}. Additionally, there is no general proof that the Wald entropy satisfies the second law of thermodynamics. There have been various attempts at designing a proof for the second law that will be valid for arbitrary higher derivative theories of gravity \cite{Iyer:1994ys, Jacobson:1993xs, Jacobson:1993vj,  Jacobson:1995uq, Sarkar:2013swa, Bhattacharjee:2015yaa, Wall:2015raa, Bhattacharjee:2015qaa, Wall:2011hj}\footnote{See the recent reviews \cite{Wall:2018ydq, Sarkar:2019xfd} and the references therein for a detailed discussion on black hole thermodynamics for higher curvature theory of gravity.}. Also, recently the construction of an entropy current in such theories was studied \cite{Bhattacharya:2019qal, Bhattacharyya:2021jhr,Bhattacharyya:2022njk}, following the work of \cite{Wall:2015raa}.

However, in this paper, our aim is to focus on the zeroth law of black hole thermodynamics in higher derivative theories of gravity. As in an ordinary thermodynamic system, the zeroth law for black hole mechanics is a characteristic signature of stationary or equilibrium black hole configurations. Stationary black holes have a space-time metric that admits a null hypersurface known as the Killing horizon, where a Killing vector becomes null. Using the fact that event horizons for stationary black holes are Killing horizons (due to the rigidity theorems), and also that the temperature of the black hole is given by the surface gravity for such stationary black hole metrics, one can make a precise statement of zeroth law as follows: \emph{the surface gravity of a stationary black hole is constant over the entire event horizon}. This statement has been proven for two derivative theories of gravity, with an additional assumption of the dominant energy condition for the matter stress tensor \cite{Bardeen:1973gs}, by analyzing the equations of motion in general relativity. Alternative proofs have been constructed \cite{Zerothwald,Xie:2021bur},  without any use of the equations of motion of the theory but assuming extra symmetries of the space-time. If we do not use any additional symmetry of the black hole space-time as mentioned above, it is an interesting question to ask if one can extend the proof of the zeroth law to theories of gravity beyond general relativity. Recently in \cite{Ghosh:2020dkk} such proof was given for stationary black hole solutions in Gauss-Bonnet and Lovelock theories of gravity by modifying and improving upon a previously reported negative result in such theories \cite{Sarkar:2012wy}; also see \cite{Sang:2021rla,Dey:2021rke} for a similar result.
As these results were worked out for particular models of higher derivative theories of gravity, to the best of our knowledge, a similar result is not yet known for arbitrary diffeomorphism invariant theories of gravity. In this paper, we address this particular question and find that the answer to this is in the affirmative: \emph{we have been able to construct a proof for the zeroth law in arbitrary diffeomorphism invariant theories of gravity where the higher derivative terms in the Lagrangian are added as a correction to the leading two derivative theory of general relativity}. 

An important assumption in our construction is the fact that it applies to theories where all higher derivative terms, appearing in the Lagrangian associated with a coupling parameter $\alpha$, are treated as corrections to a leading two derivative theory of gravity, namely Einstein's general relativity. In operational terms, this means that for theories that we consider, a smooth limit of taking the higher derivative coupling $\alpha \rightarrow 0$ exists, and in that limit, we recover the general relativity as the leading candidate theory. This, in particular, enables us to obtain stationary black hole metrics as solutions in arbitrary higher derivative theories of gravity as they can be constructed perturbatively around some known stationary black hole solutions in two derivative general relativity when $\alpha=0$. It is, however, important to highlight that our proof only requires the existence of this perturbative higher derivative coupling $\alpha$. However, it is valid for all orders in this $\alpha$-expansion and is also valid for any number of higher derivative coupling. 

Let us now mention some of the salient features of the technical tools that we have used in constructing our proof. We will be very brief here, and all of these issues will be discussed in great detail in the subsequent sections. Firstly, we will work with a particular choice for the metric of stationary black holes. This does not lose any generality as one can always make these gauge choices for any stationary black holes. For our analysis, we will focus on Killing horizons where a Killing vector becomes null on the co-dimension one null hypersurface \footnote{Sometimes, we also call it the event horizon of the black hole. However, one needs to account for various global issues in the form of rigidity theorems to ensure that the local definition of a Killing horizon can be associated with the global concept of an event horizon. It is known to be true for general relativity, but it is still an open question to prove rigidity theorems beyond general relativity. Therefore, to be precise, we will actually be working with the Killing horizon in this paper}. Further, we will associate the constancy of surface gravity on the horizon with specific components of the equations of motion. In other words, the off-shell structure of some specific components of the equations of motion, when evaluated at the horizon, will be related to (actually, be proportional to) the derivative of the surface gravity with respect to the coordinates on the horizon. Once we can establish this, the zeroth law would follow automatically by equating these components of equations of motion to zero. 
However, we must point out that we do not explicitly use equations of motion in our analysis apart from this last step. Also, the main point here is to establish the following fact - \emph{it is always possible to express the off-shell structure of a particular component of the equations of motion in arbitrary higher derivative theories of gravity (with the assumption of them augmenting the leading two derivative theory) in a form such that they get related to the derivative of surface gravity with respect to coordinates tangent to the horizon} - this is the main result of our analysis in this paper. 

As mentioned before, we organize our calculations in a perturbative expansion in the higher derivative coupling. Within such a perturbative framework, we will use the method of induction to prove that a particular component of the equation of motion has the desired off-shell structure at arbitrary order. We first argue that at the leading order, i.e., when $\alpha =0$, the equations of motion are indeed of the form expected, as this is just reviewing the proof of zeroth law known in the literature. Next, we assume that the proof works at an arbitrary order in the $\alpha$-expansion, say at $\mathcal{O} (\alpha^{m})$. Then we show that the proof will also work at the next order $\mathcal{O} (\alpha^{m+1})$. Therefore, following the method of induction, we can conclude that the proof will work up to any arbitrary higher-order in the $\alpha$-expansion. 

In establishing our result, a crucial input used a residual gauge invariance for our choice of the metric, named the boost symmetry. This boost symmetry is the consequence of a Killing isometry for stationary black holes, and the Killing horizon is mapped to itself under the flow generated by this boost transformation. Any covariant tensor, e.g., the equations of motion, will transform in a particular way under this boost transformation. This symmetry was an essential input for several recent works in the context of black hole thermodynamics. For example, in \cite{Wall:2015raa}, a proof of linearized second law for arbitrary higher derivative theories of gravity was developed using this symmetry. Also, in \cite{Bhattacharya:2019qal} and \cite{Bhattacharyya:2021jhr}, it was crucial to determine the structure of the equations of motion to construct an entropy current with non-negative divergence. In our present paper, assuming that the zeroth law is being satisfied at the order, $\mathcal{O} (\alpha^{m})$ of the $\alpha$-expansion, this boost-symmetry enables us to constrain the off-shell structure of the equations of motion at the next order $\mathcal{O} (\alpha^{m+1})$ as the desired one. 

Finally, we end this section with an overview of how the paper is structured. We begin with a description of the basic setup and an operational statement of the problem at hand in \S\ref{sec:Setup}. Here we discuss the particular choice of horizon adapted coordinates that we will use throughout this paper and present a schematic sketch of how various quantities can be organized in the perturbative expansion in the higher derivative coupling $\alpha$. In the following section \S\ref{sec:BstWght}, we present a detailed description of the boost-symmetry and the basic rules following as a consequence of this, in connection to a stationary black hole and the zeroth law. In the next section \S\ref{sec:Strategy} we briefly discuss and summarise the basic strategy of our proof without getting involved in the technical details of it. This is followed by a technically rigorous presentation of the main proof in \S\ref{sec:Proof}. We divide this into several sub-sections, each corresponding to various steps in the analysis following a method of induction. We conclude this paper with some discussions in \S\ref{sec:disco}. Important supplementary material with various technical results is presented in the Appendices \S\ref{app:kappa}-\S\ref{app:Homogeneous}.

\section{Basic set-up and statement of the problem} \label{sec:Setup}
In this section, we start by describing the basic set-up of our analysis, and we will make a precise statement of the problem using that.  

We are considering any arbitrary higher curvature theory of gravity without any matter couplings\footnote{For a proof of the zeroth law in Einstein gravity, matter couplings can be introduced with the assumption of dominant energy condition. We expect it would be straightforward to include matter couplings (even non-minimal ones) in our set-up barring some subtleties with the definition of temperature as discussed in \cite{Hajian:2020dcq}. However, we do not discuss them here.} in $d$ space-time dimensions with coordinates denoted by $x^\mu$. Following \cite{Iyer:1994ys}, the requirement of diffemorphism invariance restricts the Lagrangian for such theories to be of the following form
\begin{equation}
\mathcal{L} = \mathcal{L} (g_{\mu\nu}, \, R_{\mu\nu\alpha\beta}, \, D_\sigma R_{\mu\nu\alpha\beta}, \cdots )
\end{equation}
However, for our analysis in this paper, we will work with theories such that the gravity action has the following form
\begin{equation}\label{eq:lag}
I=\frac{1}{4\pi}\int d^dx\sqrt{-g}\left(R+\sum_{m=1}^\infty\alpha^m \, {\cal L}_{2m+2}\right)
\end{equation}
where the higher derivative couplings in the theory are denoted by the parameter $\alpha$. The other parameter present in the Lagrangian (i.e. $m$) counts the order of derivatives on the metric tensor (i.e. $g_{\mu\nu}$), the field variable in our theory. Therefore, it should be clear that ${\cal L}_{2m+2}$ is the $(2m+2)$-th order higher derivative term in the Lagrangian involving $(2m+2)$-derivatives acting on $g_{\mu\nu}$. The leading term, i.e. $m=0$, gives us the standard Einstein-Hilbert Lagrangian for general relativity. It is important to mention that, apart from having $(2m+2)$ number of derivatives on $g_{\mu\nu}$, $\mathcal{L}_{2m+2}$ has no other restrictions and is, therefore, completely arbitrary. 

Ideally, all such higher derivative terms can, in principle, appear in the Lagrangian with different numerical coefficients. Hence, one should allow for different coupling constants for each of them in different order of the parameter $m$. Even within one same order of $m$-th derivative coupling, different possible terms can appear with different coupling coefficients but with the same dimensionality \footnote{For example, let us consider the two terms at $\mathcal{O}(\alpha^2)$ in the Lagrangian: $R_{\mu}^{\nu}R_{\nu}^{\rho}R_{\rho}^{\mu}$ and $D_{\mu}R_{\nu\rho}D^\mu R^{\nu\rho}$. Both of them have six derivatives and hence can appear in the Lagrangian in the following way
$$
\alpha^2 \mathcal{L}_{6} \sim \alpha^2 (c_1 \, R_{\mu}^{\nu}R_{\nu}^{\rho}R_{\rho}^{\mu}+ c_2 \, D_{\mu}R_{\nu\rho}D^\mu R^{\nu\rho}) \, , 
$$
where $c_1$ and $c_2$ are two different but $\mathcal{O}(1)$ coefficients.}. However, as we will see in the later parts of our analysis, the only important thing for us is to have the Einstein-Hilbert term as the leading contribution in a limiting sense when the higher derivative couplings are taken to be small.
In other words, all we need is to have theories with arbitrary higher curvature terms in the Lagrangian, but any higher derivative couplings can be taken to zero in a smooth limit, leaving us with two derivative classical general relativity as the most significant one. Therefore, without any loss of generality, we collectively denote every possible higher derivative coupling by $\alpha^m$ for $m=1,\, 2,\, \cdots$, with a specific number of derivatives on $g_{\mu\nu}$ determined by the corresponding value of $m$. We will treat $\alpha$ as a small parameter allowing ourselves to perform a perturbative expansion in it. However, our analysis will be valid for arbitrary higher-order in that $\alpha$ expansion, as we have already mentioned before.

As we have described, we will be working in a perturbative expansion in the parameter $\alpha$; it is obvious that the equations of motion (EoM) will have the following structure, 
\begin{equation}\label{eq:Eom}
E_{\mu\nu}=E^{(0)}_{\mu\nu}+\alpha \, E^{(1)}_{\mu\nu}+\alpha^2\,  E^{(2)}_{\mu\nu}+ \, \cdots \, ,
\end{equation}
where, $E^{(0)}_{\mu\nu}=R_{\mu\nu}-\frac{1}{2}g_{\mu\nu}R$, is the EoM coming from Einstein's general relativity. 

Next, we would like to comment on another essential ingredient in setting up our analysis related to obtaining stationary black hole solutions in arbitrary higher derivative theories of gravity. For purposes of the arguments presented in this paper, we do not need to know the exact form of the stationary black hole metric as a solution to the equations of motion. However, we assume that such solutions must exist in the higher derivative theory of gravity that we are considering. One should, quite naturally, be able to construct such solutions \cite{Cai_2002, Ma_2021} within our setup of perturbative expansion in $\alpha$, the coupling of the higher derivative terms in our theory. 

Let us suppose we start with a given stationary black hole solution, denoted by $g^{(bh)}_{\mu\nu}$, in the leading order theory in $\alpha$ expansion, which is Einstein's general relativity. It is obvious that the stationary $g_{\mu\nu} = g^{(bh)}_{\mu\nu}$ solves the equation of motion $E^{(0)}_{\mu\nu}=0$. As a consequence of this, $g^{(bh)}_{\mu\nu}$ will have a Killing horizon - a null hypersurface generated by a global Killing vector field which we will denote by $\partial_\tau$. By definition, $\partial_\tau$ will be a null geodesic on the horizon, and all the metric components in $g^{(bh)}_{\mu\nu}$ will be independent of the $\tau$ coordinate. In the following paragraphs, we will make this more precise.

We will be working with a particular horizon adapted set of the space-time coordinates along with a particular gauge choice for the metric $g^{(bh)}_{\mu\nu}$. In a $d$-dimensional space-time we can always choose a coordinate system  $x^\mu=\{\tau,\rho,x^i\}$, where $i=1,...,d-2$, so that the stationary metric $g^{(bh)}_{\mu\nu}$ takes the following form
\begin{equation}\label{eq:metric}
\begin{split}
ds^2=g^{(bh)}_{\mu\nu} \, dx^\mu dx^\nu
=2  \, d\tau \, d\rho-\rho \,  X(\rho,x^i) \, d\tau^2+2 \, \rho \,  \omega_i(\rho,x^i)  \, d\tau  \, d x^i+h_{ij}(\rho,x^i)  \, dx^i  \, dx^j \, .
\end{split}
\end{equation}
Let us briefly justify the gauge choice for the metric in eq.\eqref{eq:metric} (see section-(2.1) and Appendix-A of \cite{Bhattacharyya:2016xfs} for the details). The coordinates $\{\tau, \, x^i\}$ span the co-dimension one horizon which lies on $\rho=0$. We should also note that $\rho =0$ is a null hypersurface, for which the null generators are taken to be the vector $\xi=\partial_\tau$. By construction, this is normal to itself and the other spatial generators ($\partial_i$) of the horizon. At constant values of the coordinates $x^i$, the parameter $\tau$ runs along one null generator, whereas, for a constant value of $\tau$, the coordinates $x^i$ parametrizes different null generators on the horizon.  The coordinate $\tau$ is not necessarily affinely parametrized. To describe the geometry in the vicinity of a null hypersurface, we need two null normals to it. Hence, apart from $\xi$, we have considered the auxiliary vector $\chi = \partial_\rho$, which is also null. This gives us the coordinate $\rho$, which parametrizes the distance away from the null horizon. The coordinate $\rho$ has been chosen to be affinely parametrized, and the inner products: $(\partial_\tau , \partial_\rho) |_{\rho =0}= 1$ and $(\partial_i, \partial_\rho) |_{\rho =0}= 0$, define the angles with which the null-vector $\partial_\rho$ pierces through the horizon at $\rho=0$. 

Next, the additional requirement of stationarity should explain why the metric coefficients (the functions $X$, $\omega_i$, and $h_{ij}$) are independent of the coordinate $\tau$.  To this we note that $\xi$ is a Killing vector for the metric eq.\eqref{eq:metric}, satisfying the Killing equation $D_\mu \xi_\nu + D_\nu \xi_\mu =0$, where $D_\mu$ is the covariant derivative with respect to the full black hole metric, $g^{(bh)}_{\mu\nu}$. The norm of this Killing vector vanishes on the surface $\rho =0$. Thus, in our choice of coordinates, $\rho = 0$ hypersurface is a Killing horizon. 

The vector field $\xi^\mu$ also satisfies the geodesic equation 
\begin{equation}
\xi^\nu \, D_\nu  \xi^\mu = \kappa \, \xi^\mu \,. 
\end{equation}
Note that, the RHS of the above equation is not zero since $\tau$ is not necessarily an affine parameter. This equation could be considered as the definition of the quantity $\kappa$, which is in general a function of the coordinates $(\tau, \, x^i)$ and is called the surface gravity. It can be straightforwardly shown that the surface gravity for the black hole space-time described by the metric given in eq.\eqref{eq:metric} can be written as
\begin{equation} \label{defkappa}
\kappa=\left. \sqrt{-\frac{1}{2}\,(D_\mu \xi_\nu)\,(D^\mu \xi^\nu)} \, \right\vert_\text{horizon} \, . 
\end{equation}

The surface gravity is related to the temperature of a stationary black hole, and thus to prove the zeroth law, we must show that $\kappa$ is constant over the horizon. It means that the surface gravity is constant not only for evolutions along one null generator but also does not change across different null generators of the null horizon. In other words, we would aim to prove that, when evaluated on the horizon,
\begin{equation}
\partial_\tau \kappa =0, \quad \text{and} \quad \partial_i\kappa =0 \, .
\end{equation}

Following the definition in eq.\eqref{defkappa}, we can evaluate the surface gravity for our choice of metric eq.\eqref{eq:metric} for $\xi = \partial_\tau$, to get the following expression (see Appendix-\ref{app:kappa} for details of the calculation)
\begin{equation} \label{kappa_expr}
\kappa = \left. {1\over 2} X(\rho,x^i) \, \right \vert_{\rho=0} .
\end{equation}
It is obvious from eq.\eqref{kappa_expr} that $\kappa$ is independent of the coordinate $\tau$. Basically, since $\xi$ is a Killing vector, we trivially obtain the $\tau$ independence of $X(\rho,x^i)$, and hence $\partial_\tau \kappa =0$. Therefore, to prove the zeroth law we have to show the following on the horizon
\begin{equation}\label{eq:finalshow}
\partial_i X(\rho,x^i) \, |_{\rho=0}=0 \, .
\end{equation}

\section{Boost symmetry in the context of the zeroth law and stationarity} \label{sec:BstWght}
As we have laid down the statement of the problem in operational terms, in this section, we would like to highlight one crucial significance of the zeroth law or, equivalently, the constancy of surface gravity over the horizon. Let us remind ourselves that the zeroth law is, in a sense, one particular manifestation of stationarity for black hole solutions in our theory. It is noteworthy that for our choice of the stationary black hole metric in eq.\eqref{eq:metric} the coordinate $\tau$ runs along the null generators of the horizon but is not affinely parametrized. However, a slightly different but very useful choice of coordinate system as written below
\begin{equation} \label{eq:metricAFF}
ds^2=\tilde{g}^{(bh)}_{\mu\nu} \, dx^\mu dx^\nu
=2  \, dv \, dr-r^2 \,  X(rv, \, x^i) \, dv^2+2 \, r \,  \omega_i(rv, \, x^i)  \, dv  \, d x^i+h_{ij}(rv, \, x^i)  \, dx^i  \, dx^j \, ,
\end{equation}
also describes metric of stationary black holes with the horizon being set at $r=0$, see \cite{Wall:2015raa}, \cite{Bhattacharya:2019qal}, \cite{Bhattacharyya:2021jhr}. The crucial difference between this choice of metric in eq.\eqref{eq:metricAFF}, written in terms of the new coordinates $(r, \, v, \, x^i)$, compared to the one in eq.\eqref{eq:metric}, is the fact that the $v$ coordinate here is affinely parametrized along the null generators $\partial_v$ of the horizon. It should also be noted that, although, for the choice of metric in eq.\eqref{eq:metric} the metric coefficients are independent of the parameter $\tau$, in eq.\eqref{eq:metricAFF} the metric coefficients are functions of the coordinate $v$. However, the dependence on $v$ is not arbitrary but restricted to the product $r\,v$. The reason for this is the following, for stationary metrics, the Killing generator $\partial_\tau$ and the affinely parametrized null generators are not the same but proportional to each other, see Appendix-A of \cite{Bhattacharya:2019qal} for a detailed discussion on this. 

Let us now highlight the usefulness of writing the stationary black hole metric in the form of eq.\eqref{eq:metricAFF} with $v$ being an affine parameter. This particular choice does not fix the gauge completely and one still has some residual freedom of performing further coordinate transformation. Particularly, one can do the following scaling of the coordinates $(r, \, v)$ 
\begin{equation} \label{eq:BoostTrans}
r \rightarrow \lambda\, r \, , \quad  \text{accompanied with} \quad v \rightarrow {v \over \lambda} \, ,
\end{equation}
where $\lambda$ is a constant parameter \footnote{Actually, one can do a more general residual coordinate transformation $v \rightarrow f_1(x^i) v + f_2(x^i), $ along with appropriate redefinition of $r$, but here we have restricted ourselves  to a subclass of it.}. It should be convincing that this transformation should leave the metric invariant, since the metric functions depend on the coordinates $(r, \, v)$ only through their product. This is called the boost transformation and due to this the stationary black hole configurations are said to enjoy a boost symmetry, see \cite{Bhattacharya:2019qal}, \cite{Bhattacharyya:2021jhr} for details. 

Alternatively, we can also explain the boost symmetry, that we described above, in the following way. In the coordinate system $\{r, \, v, \, x\}$, a stationary black hole solution, as written in eq.\eqref{eq:metricAFF}, has a Killing vector 
\begin{equation} \label{killingvector}
\xi = \xi^\mu\partial_\mu =( v\partial_v - r\partial_r) \, .
\end{equation}
In other words, the metric eq.\eqref{eq:metricAFF} satisfies the following
\begin{equation}
\mathcal{L}_\xi g_{\mu\nu}^{bh} = 0 \, ,
\end{equation}
where $\mathcal{L}_\xi$ denotes the Lie derivative with respect to the vector $\xi$. It can be easily checked that $\xi$ is also the generator of the infinitesimal version of the boost transformation eq.\eqref{eq:BoostTrans}.   

As a consequence of this Killing symmetry, we can also confirm that the Lie derivative of any arbitrary covariant tensor constructed out of the metric should also vanish. The boost-symmetry is extremely useful in determining how any general tensor quantity built out of the metric coefficients or various derivatives of them,  would transform under the aforementioned boost-transformation. In particular, any covariant tensor, say $\mathcal{B}$, with all components lowered, would transform in the following way 
\begin{equation} \label{defboostwght}
\mathcal{B} \rightarrow \widetilde{\mathcal{B}} = \lambda^w \, \mathcal{B}, \quad \text{under} \quad \left( r \rightarrow \tilde{r} = \lambda\, r \, , \, v \rightarrow \tilde{v} =  {v \over \lambda} \right) \, 
\end{equation}
so that we define the boost-weight of $\mathcal{B}$ to be given by $w$. Alternatively, we can also show that the boost-weight of any covariant tensor would be given by the number of excess lower $v$-indices over the lower $r$-indices, see \cite{Bhattacharyya:2021jhr} and Appendix-\ref{app:ReviewBoostWeight} for a justification in favor of this. 

Let us mention one important result that follows from the set up of boost-symmetry discussed above, any quantity with positive boost-weight will always vanish when computed using metric corresponding to a stationary configurations and evaluated on the Killing horizon. In order to explain this statement, let us first note that, from the definition of boost weight given in eq.\eqref{defboostwght} it can be argued that the metric functions, $X(rv, \, x^i), \, \omega_i(rv, \, x^i) , \, h_{ij}(rv, \, x^i)$ appearing in eq.\eqref{eq:metricAFF}, are all boost invariant objects.  Additionally the derivatives $\partial_v$ and $\partial_r$ have boost weights given by $+1$ and $-1$ respectively, 
\begin{equation}
\partial_v \rightarrow \lambda \, \partial_v \, , \quad \text{and} \quad \partial_r \rightarrow \lambda^{-1} \, \partial_r \, . 
\end{equation}
Therefore, any covariant tensor, say $\mathcal{B} (rv, \, x^i)$, with positive boost weight can generically be written as 
\begin{equation}
\mathcal{B} (rv, \, x^i) \sim (\partial_r)^{m_r} (\partial_v)^{m_v} \, \widetilde{\mathcal{B}}(rv, \, x^i) \, , \quad \text{with} \quad m_v > m_r \, ,
\end{equation}
where $\widetilde{\mathcal{B}}(rv, \, x^i)$ can include derivatives with respect to the spatial coordinates, but not any $\partial_v$ or $\partial_r$. The functional dependence of $\mathcal{B}(rv, \, x^i)$ or $\widetilde{\mathcal{B}}(rv, \, x^i)$ on the product of $rv$ signifies that they are evaluated on stationary configurations. Because of $m_v > m_r $, $\mathcal{B}$ has positive boost weight equal to $(m_v-m_r)$. Now, it is easy to convince ourselves that whenever one operates $(\partial_r)^{m_r} (\partial_v)^{m_v}$ on $\widetilde{\mathcal{B}}(rv, \, x^i)$, or in that case any function of the product $rv$, $(m_v-m_r)$ factors of $r$ will be obtained, and hence it will vanish when we further evaluate this on the horizon $r=0$. This will also be very crucially used in our present paper.

In \cite{Bhattacharya:2019qal} and \cite{Bhattacharyya:2021jhr}, this particular boost symmetry was used to construct a local entropy current with non-negative divergence on the horizon of a dynamically perturbed stationary black hole in an arbitrary diffeomorphism invariant theory of gravity. In order to study non-stationary dynamical processes, this boost-symmetry is broken slightly by some matter source hitting the stationary black hole space-time. One can organize the dynamics in a perturbative expansion around the initial stationary configuration in the small amplitude of the external matter disturbance. Up to linearized order in the expansion in this amplitude expansion, the $vv$-component of the equations of motion (EoM) in any diffeomorphism invariant theory of gravity  attains a universal structure as given below
\begin{equation}
E_{vv} \sim \partial_v \left(\partial_v J^v + \nabla_i J^i \right) + \text{quadratic fluctuations} \, , 
\end{equation}
where the quantity $J^v$ represents local entropy density, reproducing the Wald entropy expression upon taking the stationary limit. On the other hand, the spatial components $J^i$ signify the spatial flow of entropy on constant $v$-slices of the horizon. Using this result obtained in general gravity theories, one further needs to use the null energy condition for the stress-energy tensor coming from the matter sector to construct a proof for the local version of a second law.

It was, therefore, indeed essential for the analysis in \cite{Bhattacharyya:2021jhr} to have the stationary metric written in the form given in eq.\eqref{eq:metricAFF}. In this section, we will argue that if the zeroth law is satisfied one can perform a coordinate transformation that changes the space-time metric from eq.\eqref{eq:metric} to eq.\eqref{eq:metricAFF}. Although this was implicit in the calculations in Appendix-A of \cite{Bhattacharyya:2021jhr}, here we would like to make it very explicit. 

Once zeroth law is satisfied, we get the surface gravity constant over the horizon. Therefore, we should be able to solve eq.\eqref{eq:finalshow} and obtain the following general solution for the metric coefficient function $X(\rho, \, x^i)$
\begin{equation}
X(\rho, \, x^i) = c_1 + \rho \, f(\rho, \, x^i) \, ,
\end{equation}
where $c_1$ is an integration constant and $f(\rho, \, x^i)$ is some arbitrary function of $(\rho, \, x^i)$. Also, note that in order to satisfy eq.\eqref{kappa_expr}, the constant $c_1$ gets fixed as $c_1 = 2 \,  \kappa$. We can substitute this in eq.\eqref{eq:metric} to obtain
\begin{equation}\label{eq:metric2}
\begin{split}
ds^2=2  \, d\tau \, d\rho-\rho \,  \left(c_1 + \rho \, f(\rho, \, x^i)\right)\, d\tau^2+2 \, \rho \,  \omega_i(\rho,x^i)  \, d\tau  \, d x^i+h_{ij}(\rho,x^i)  \, dx^i  \, dx^j \, .
\end{split}
\end{equation}
Next we perform the following coordinate transformation from the coordinates $\{\rho, \, \tau, \, x^i \}$ to $\{r, \, v, \, x^i \}$ given by 
\begin{equation}\label{coordtrans1}
\tau \rightarrow v=\frac{2}{c_1}\exp\left(\frac{c_1}{2}\tau\right), \quad \text{and}, \quad \rho \rightarrow r=\rho \exp\left(-\frac{c_1}{2}\tau\right)
\end{equation}
to arrive at 
\begin{equation}\label{eq:station}
ds^2=2 \, dv\, dr-r^2 \, f(c_1\, r \, v/2, \, x^i) \, dv^2+r \, \omega_i(c_1 \, r \, v/2, \, x^i) \, dv \, dx^i+h_{ij}(c_1 \, r \, v/2, \, x^i) \, dx^i dx^j \, ,
\end{equation}
which is of the form eq.\eqref{eq:metricAFF}. 
Note that the horizon stays at $r=0$ in the new coordinates, and, also, the fact that $c_1 = 2 \kappa$ is a constant was crucially used while performing this coordinate transformation. Once written in this coordinate system, we can straightforwardly use the consequences of boost-symmetry that the metric in this form enjoys. 

Finally, before we end this section, let us make one comment on how these results that one derives using boost invariance of a stationary black hole expressed in the coordinates as in eq.\eqref{eq:station}, would be helpful in the later sections of this paper as we aim to prove zeroth law. This may seem puzzling since, to derive these results, we have already used the zeroth law itself. However, as we will explain later, we will follow a methodology for the proof of zeroth law by organizing our calculations as a perturbative correction in the higher derivative coupling $\alpha$ correcting the leading order two derivative theory of general relativity. In that perturbative set-up, we will construct the proof by using a method of induction. More precisely, with the assumption that at $n$-th order in the $\alpha$-expansion, our construction validates the zeroth law, we will aim to extend the proof to $n+1$-th order. Therefore, while working at $n+1$-th order, the truncated and corrected metric till the previous $n$-th order could be brought to the form as in eq.\eqref{eq:station} and thus would satisfy boost-invariance under the transformation given in eq.\eqref{eq:BoostTrans}. Consequently, when evaluated on the metric corrected and truncated up to $n$-th order in $\alpha$-expansion, any covariant tensor would transform with a particular boost-weight entirely determined by its index structure alone. For our case, using these concepts, we will see that the $(vi)$-component of EoM, i.e., $E_{vi}$, would have a boost weight equal to $+1$ and would thus vanish for stationary black hole configurations.

\section{Brief outline of the strategy} \label{sec:Strategy}
In this section, our goal is to present a broad outline of the strategy of the proof without getting into operational details. Following this, in the next section, we will construct a technically rigorous proof. 

The crucial ingredient in our strategy to prove the zeroth law will be to argue that eq.\eqref{eq:finalshow} follows from the vanishing of a particular component of the equations of motion (EoM). More precisely, we will explicitly show that the LHS of eq.\eqref{eq:finalshow} must be 
expressed in terms of the $\{\tau \, i\}$-component of the EoM
\begin{equation} \label{eq:maingoal}
E_{\tau i}  \, |_{\rho=0}\sim \partial_i X(\rho,x^i)  \, |_{\rho=0}\, ,
\end{equation}
upto numerical factors, where, following eq.\eqref{eq:Eom}, $E_{\tau i}$ must include contributions from all higher derivative terms present in the Lagrangian of the theory in addition to the leading Einstein-Hilbert term. Next, we should use $E_{\tau i}=0$, as the stationary black hole space-times must solve the full EoM's. It is clear that eq.\eqref{eq:finalshow} follows immediately.  

With the explanations so far, let us summarise the main goal that we will pursue in the rest of this paper: \emph{In order to prove the zeroth law (or equivalently, for proof of eq.\eqref{eq:finalshow}), our primary goal would be to justify that the off-shell structure of the entire $E_{\tau i}$ (including corrections due to higher derivative coupling $\alpha$ as in eq.\eqref{eq:Eom}) reproduces eq.\eqref{eq:maingoal}. To achieve this, we must evaluate $E_{\tau i}$ for a stationary black hole solution obtained by treating the higher derivative coupling $\alpha$ perturbatively in an expansion around a stationary black hole solution of the leading two-derivative theory.}

Before we proceed further with describing our strategy, let us take a small detour to highlight the significance of this particular component of EoM's, $E_{\tau i}$, in justifying the zeroth law. On general grounds, looking at eq.\eqref{eq:finalshow}, we should expect that $\partial_i X(\rho, x^i)\, |_{\rho=0}$ should get related to some components of $E_{\mu\nu}$ in order to be vanishing when evaluated on on-shell configurations. The index structure then suggests that one of the two indices in $E_{\mu\nu}$ must be the spatial indices (say, $\mu =i$), leaving the other one to be either $\nu = \tau$, or, $\nu = \rho$. If we now focus on Einstein's gravity and compute the $(\tau, \, i)$ component of the Einstein tensor, which is the EoM, we can immediately check that eq.\eqref{eq:maingoal} is reproduced. Hence, the zeroth law is proved for the two derivative theory. In our set-up, we treat arbitrary theories of gravity as perturbative corrections in the higher derivative coupling $\alpha$ to a leading two derivative theory. Hence, it is expected that even in such general theories, we must look into the off-shell structure of $E_{\tau i}$ to justify eq.\eqref{eq:finalshow}. 

The arguments presented in the previous paragraph may appear to be heuristic. However, a more rigorous justification can be devised to support the following statement: for a proof of the zeroth law, one should investigate the off-shell structure of $(\tau \, i)$-component of EoM. This has already been noted in the literature. It is possible to show that (see \cite{Bardeen:1973gs} for proof)
\begin{equation} \label{DiKappaRel}
\begin{split}
e^{\mu}_i D_{\mu} \, \kappa &= - R_{\mu\nu} \, \xi^\mu\,  e^{\nu}_i \, ,
\end{split}
\end{equation}
where $e^{\mu}_i$ are the space-like tangent vectors to the horizon at $\rho=0$. Most significantly, we should note that to derive eq.\eqref{DiKappaRel} one does not need to use any EoM, and hence this is valid universally in any theory of gravity. For two derivative Einstein gravity, once we use EoM, the RHS in eq.\eqref{DiKappaRel} gets related to components of the stress-energy tensor, $T_{\mu\nu}$, coming from the matter sector coupled to gravity, if any. One can further use the dominant energy condition for the stress-energy tensor, and consequently, the RHS in eq.\eqref{DiKappaRel} vanishes, proving the zeroth law. However, once we focus on higher derivative theories of gravity, in this process of substituting $R_{\mu\nu}$ in terms of $T_{\mu\nu}$ we get extra contributions in the EoM due to the higher derivative terms in the Lagrangian
\begin{equation}
E^{(0)}_{\mu\nu}+\alpha \, E^\text{HD}_{\mu\nu} = T_{\mu\nu} \, ,
\end{equation}
where $E^{(0)}_{\mu\nu}=R_{\mu\nu} -(1/2) \, g_{\mu\nu} \, R$ is the Einstein tensor, and $E^\text{HD}_{\mu\nu}$ is the EoM coming from the higher derivative terms in the theory along with the coupling $\alpha$. With this, we are convinced that in order to prove that the RHS of eq.\eqref{DiKappaRel} vanishes, we must need to investigate the off-shell structure of $E^\text{HD}_{\mu\nu}\, \xi^\mu\,  e^{\nu}_i$, which, in our chosen coordinate system, is precisely the $E^\text{HD}_{\tau i}$. 

With the set-up that we have discussed so far, we are now in a position to give a schematic overview of the operational strategy that will be followed to argue that $E_{\tau i}$ is indeed of the form given in eq.\eqref{eq:maingoal}. The main idea will be to organize the analysis in a perturbative expansion in higher derivative coupling $\alpha$ around the leading two derivative theory. The EoM has already been written in eq.\eqref{eq:Eom} in order by order expansion in $\alpha$. In order to investigate the off-shell structure of $E_{\mu\nu}$ constructed out of the space-time metric $g_{\mu\nu}$ and derivatives acting on it, we would also need to take a similar ansatz for $g_{\mu\nu}$ expanded in powers of $\alpha$
\begin{equation} \label{MetExpnsn}
g^{(bh)}_{\mu\nu} = g^{(0)}_{\mu\nu} + \alpha \, g^{(1)}_{\mu\nu} +\alpha^2 \, g^{(2)}_{\mu\nu} + \cdots \, , 
\end{equation}
where the superscript in $g^{(n)}_{\mu\nu}$ signifies that it corresponds to the $n$-th order in the expansion of $\alpha$. We would demand that this ansatz for $g^{(bh)}_{\mu\nu}$ solves the EoM given in eq.\eqref{eq:Eom} order by order in the expansion of $\alpha$. This, in turn, would allow us to justify that eq.\eqref{eq:maingoal} is indeed true up to all orders in the perturbative expansion in $\alpha$. 

We will follow the method of induction to establish the desired off-shell structure of  $E_{\tau i}$ up to arbitrary order in the $\alpha$-expansion. First, we would show that it is indeed the case at the leading order with $\alpha = 0$ for two-derivative Einstein gravity. Then we will extend this to arbitrary order of $\mathcal{O}(\alpha^{m+1})$ in the $\alpha$-perturbative expansion, assuming that things do work out till the previous order of $\mathcal{O}(\alpha^{m})$. In this process, we will need to know specifics about the generic structure of $E_{\mu\nu}$ in an arbitrary order of the perturbation. To evaluate $E_{\mu\nu}$, we need to substitute for $g^{(bh)}_{\mu\nu}$, given in eq.\eqref{MetExpnsn},  in eq.\eqref{eq:Eom} and isolate the terms contributing at $\mathcal{O}(\alpha^{m+1})$.  We will see that at this order (i.e. at $\mathcal{O}(\alpha^{m+1})$), $E_{\mu\nu}$ can be partitioned into two types of terms. 
The first type being the zeroth order EoM $E^{(0)}_{\mu\nu}$ evaluated on $g^{(0)}_{\mu\nu}+\alpha^{m+1}g^{(m+1)}_{\mu\nu}$, where we will treat $\alpha^{m+1}g^{(m+1)}_{\mu\nu}$ as linearized perturbation around $g^{(0)}_{\mu\nu}$. The second type of terms involve the coefficient of $\alpha^{m+1}$ in the full EoM evaluated on the metric corrected till the previous order, i.e. till $g^{(m)}_{\mu\nu}$. Schematically, this looks as the following 
\begin{equation} \label{EoM_at_n+1}
\text{At} \, \mathcal{O}(\alpha^{m+1}): \quad E^{(0)}_{\mu\nu}[g^{(0)}_{\mu\nu}+\alpha^{m+1} \, g^{(m+1)}_{\mu\nu}] + E_{\mu\nu}[ g^{(0)}_{\mu\nu} + \alpha \, g^{(1)}_{\mu\nu} +\alpha^2 \, g^{(2)}_{\mu\nu} + \cdots + \alpha^{m} \, g^{(m)}_{\mu\nu}] = \mathcal{O}(\alpha^{m+2}) \, ,
\end{equation} 
where for the second term on the LHS we should truncate it to $\mathcal{O} (\alpha^{m+1})$. 

The first term on the LHS of eq.\eqref{EoM_at_n+1} has an universal structure as it is basically the Einstein's tensor linearized around $g^{(0)}_{\mu\nu}$ for a small perturbation given by $g^{(m+1)}_{\mu\nu}$.

To treat the second term on the LHS in eq.\eqref{EoM_at_n+1}, however, we have to be more careful. Since this term has no universal structure like the first one, a further non-trivial argument must be invoked. We should take note of the fact that this second term is an arbitrary covariant tensor of rank two, but most importantly, built out of metric coefficients truncated at $\mathcal{O}(\alpha^{m})$. When we are looking at the order ${\cal O}(\alpha^{m+1})$, we will assume that eq.\eqref{eq:maingoal} and eq.\eqref{eq:finalshow} have been satisfied till the order of $\mathcal{O}(\alpha^{m})$. This, in turn, enables us to ascertain that the surface gravity, $\kappa$, computed with the corrected metric till $\mathcal{O}(\alpha^{m})$, will be constant over the horizon. As a consequence of this, we know that $\partial_i X(\rho, \, x^i) |_{\rho=0} =0$ up to $\mathcal{O}(\alpha^{m})$, and hence we can use the coordinate transformation eq.\eqref{coordtrans2} to write the metric in terms of coordinates $(r, \, v, \, x^i)$, as in eq.\eqref{eq:station1}. As we have discussed before, once we have succeeded in writing the space-time metric in this new coordinate system, we can use the boost-symmetry. Consequently, we would now be able to assign boost weights to various covariant tensors just by counting the difference in lower $v$ and $r$ components. By standard rule of how tensors should transform under coordinate transformation one can see that $E_{\tau i}$ in the original $(\rho, \, \tau, \, i)$-coordinates will be proportional to $\tilde{E}_{vi}$ - the $(v i)$-component of the new EoM in the new coordinates. Finally, we note that $\tilde{E}_{vi}$ has boost-weight$ = +1$, and hence should vanish when computed for stationary configurations. Therefore we conclude that the second term on the LHS of eq.\eqref{EoM_at_n+1} will not contribute at $\mathcal{O}(\alpha^{m+1})$. With this, we will be able to show that $E_{\tau i}$ has indeed the form mentioned in eq.\eqref{eq:maingoal} and, thereby, we will be able to complete the proof of the zeroth law.

\section{Constructing the proof for the zeroth law} \label{sec:Proof}
This section will construct the proof for the zeroth law with all details.
The discussions in this section will be divided into several sub-sections. Following the outline of our strategy presented in \S\ref{sec:Strategy}, we would start with understanding the general structure of the equations of motion (EoM) order by order in an expansion in the higher derivative coupling $\alpha$. This would be followed by explicitly working out the leading order term in this expansion, which is actually the Einstein tensor coming from the Einstein-Hilbert Lagrangian. We would review how zeroth law is satisfied at this leading order. Next, we would extend this procedure to an arbitrary higher-order in the $\alpha$-expansion adopting a method of induction. Finally, we will also explicitly see how the application of boost symmetry for stationary metric corrected up to a particular order of $\alpha$-expansion helps us determine the $(\tau i)$-components of the EoM to the following order.

\subsection{General structure for the equations of motion in $\alpha$-expansion}\label{subsec:structure}
As we have already mentioned before in \S\ref{sec:Setup}, the key assumption in our working principle is that we could solve the EoM perturbatively in an expansion in the higher derivative coupling $\alpha$. Thereby, the EoM has a structure given in eq.\eqref{eq:Eom}, which we rewrite here for convenience
\begin{equation}\label{eq:Eom1}
E_{\mu\nu}=E^{(0)}_{\mu\nu}+\alpha \, E^{(1)}_{\mu\nu}+\alpha^2\,  E^{(2)}_{\mu\nu}+ \, \cdots \, ,
\end{equation}
where $E^{(0)}_{\mu\nu} $ is the Einstein tensor - the EoM in the two derivative theory of gravity. Also, $\alpha^k E^{(k)}_{\mu\nu}$, for $k\geq1$, are all higher derivative corrections to the EoM. They depend on the details of the theory. 
We have also noticed in \S\ref{sec:Strategy}, that as a consequence, the metric will also admit a similar expansion, given in eq.\eqref{MetExpnsn}. Here we present that as well
\begin{equation} \label{MetExpnsn1}
g^{(bh)}_{\mu\nu} = g^{(0)}_{\mu\nu} + \alpha \, g^{(1)}_{\mu\nu} +\alpha^2 \, g^{(2)}_{\mu\nu} + \cdots \, .
\end{equation}
In our choice of coordinates this will lead to an expansion of the metric components $X$, $\omega_i$ and $h_{ij}$ as given below
\begin{equation}\label{eq:metexpansion}
\begin{split}
X(\rho,x^i)&=X^{(0)}(\rho,x^i)+\alpha \, X^{(1)}(\rho,x^i)+\alpha^2 \,  X^{(2)}(\rho,x^i)+\cdots  \, ,\\
\omega_i(\rho,x^i)&=\omega_i^{(0)}(\rho,x^i)+\alpha \, \omega_i^{(1)}(\rho,x^i)+\alpha^2 \,  \omega_i^{(2)}(\rho,x^i)+\cdots \,  ,\\
h_{ij}(\rho,x^i)&=h^{(0)}_{ij}(\rho,x^i)+\alpha \,  h^{(1)}_{ij}(\rho,x^i)+\alpha^2 \,  h^{(2)}_{ij}(\rho,x^i)+\cdots \, .
\end{split}
\end{equation}

We are viewing the EoM in eq.\eqref{eq:Eom1} to be evaluated on the metric $g^{(bh)}_{\mu\nu}$ in eq.\eqref{MetExpnsn1}, and the corresponding structures should be analysed order by order in the $\alpha$-expansion. Let us now see what we can learn at the very leading order, i.e. at $\mathcal{O}(\alpha^0)$. At this order very leading order the metric functions $X^{(0)}(\rho,x^i), \, \omega_i^{(0)}(\rho,x^i)$ and $h^{(0)}_{ij}(\rho,x^i)$ should be exact solutions of the zeroth order equation (i.e. the Einstein's equation for two derivative theory of gravity)
\begin{equation}
E^{(0)}_{\mu\nu} [g^{(0)}_{\mu\nu}] = 0
\end{equation}
Therefore, we should be viewing this as a differential equation for $g^{(0)}_{\mu\nu}$, which is the unknown variable at leading order, and by solving this, we would be able to fix $g^{(0)}_{\mu\nu}$.

Now, let us suppose that we want to solve the EoM $E_{\mu\nu} =0$ upto the first sub-leading order (i.e., upto order ${\cal O}(\alpha^1)$). We have already determined $g^{(0)}_{\mu\nu}$ while working at the previous order ${\cal O}(\alpha^0)$. At this order of ${\mathcal O}(\alpha^1)$, we realise that $g^{(1)}_{\mu\nu}$ (or the metric functions  $X^{(1)}(\rho,x^i), \, \omega_i^{(1)}(\rho,x^i)$ and $h^{(1)}_{ij}(\rho,x^i)$) are the unknowns. To find out the relevant part of EoM from eq.\eqref{eq:Eom1} at this order, we will basically have to evaluate the tensor $E_{\mu\nu}$ on the metric $g_{\mu\nu}^{(bh)}$, neglecting all terms proportional to quadratic or higher powers of $\alpha$. In other words, it is obvious that $E^{(n)}_{\mu\nu}$ and $g^{(n)}_{\mu\nu}$ for every $n \geq 2$ are negligible at this order of  ${\cal O}(\alpha^1)$. As a result, the differential equation for the unknowns will have the following structure
\begin{equation}\label{eq:2ndorder}
E_{\mu\nu}^{(0)}\left[g^{(0)}_{\alpha\beta}+\alpha~ g^{(1)}_{\alpha\beta}\right] +\alpha~ E_{\mu\nu}^{(1)}\left[g^{(0)}_{\alpha\beta}\right] =\mathcal{O} (\alpha^2) \, ,
\end{equation}
Here, in eq.\eqref{eq:2ndorder}, the first term on the LHS is basically the Einstein tensor, linearised around $g^{(0)}_{\mu\nu}$ where $g_{\mu\nu}^{(1)}$ plays the role of the small fluctuation metric. The second term is actually not of any universal structure like Einstein tensor since the explicit form would depend on the type of higher derivative theory that we are focussing on at linear order in $\alpha$. However, for our purpose, that is not at all a problem since we just need to know that $E_{\mu\nu}^{(1)}$ is a covariant tensor of rank two (constructed out of appropriate contractions of Riemann tensors and its covariant derivatives) evaluated on the exact stationary black hole solution, i.e., $g^{(0)}_{\mu\nu}$, of the two-derivative theory of gravity. From eq.\eqref{eq:2ndorder}, it is clear that this term will act as a source term in the inhomogeneous PDE for $g^{(1)}_{\mu\nu}$. The first term, on the other hand, is homogeneous in $g^{(1)}_{\mu\nu}$ and has a known and universal structure. 

Consequently, at the next sub-leading order at ${\mathcal O}(\alpha^2)$, the unknowns are $g^{(2)}_{\mu\nu}$ (or, as usual, $X^{(2)}(\rho,x^i), \, \omega_i^{(2)}(\rho,x^i)$ and $h^{(2)}_{ij}(\rho,x^i)$), and the PDE for them should look like 
\begin{equation}\label{eq:3rdorder}
E_{\mu\nu}^{(0)}\left[g^{(0)}_{\alpha\beta}+\alpha^2 \, g^{(2)}_{\alpha\beta}\right] + E_{\mu\nu}\left[g^{(0)}_{\alpha\beta} + \alpha \, g^{(1)}_{\alpha\beta}\right] = {\mathcal O}(\alpha^3) \, ,
\end{equation}
where the first term on LHS is the homogeneous piece and the other term is the source term for  $g^{(2)}_{\mu\nu}$. The source term, again, is evaluated on metric functions $g^{(1)}_{\mu\nu}$ and $g^{(0)}_{\mu\nu}$, which are already solved in the previous iteration at ${\mathcal O}(\alpha^2)$. 

Once we have studied the EoM to the second sub-leading order in the $\alpha$-expansion, we should be able to extend this analysis to any arbitrary order in $\alpha$. Let us assume that we are currently focussing on the $(m+1)$-th order term in the expansion. We have learned that if we would like to determine the solution correctly up to order ${\mathcal O}(\alpha^{m+1})$, we have to evaluate $E_{\mu\nu}$ on $g_{\mu\nu}^{(bh)}$ neglecting all terms of order ${\mathcal O}(\alpha^{m+2})$ and higher. 
At this order, i.e. at ${\cal O}(\alpha^{m+1})$, the unknowns would be the components of $g^{(m+1)}_{\mu\nu}$ or in our choice of gauge, the metric functions $X^{(m+1)}(\rho,x^i), \, \omega_i^{(m+1)}(\rho,x^i)$ and $h^{(m+1)}_{ij}(\rho,x^i)$. Now, we would like to know what would be the structure of the EoM at this order. As it is true for any perturbative solution technique, the homogeneous part of the equation at every order has an universal structure. In this case it is the Einstein equation $E^{(0)}_{\mu\nu}$ linearized around the zeroth order black hole metric $g^{(0)}_{\mu\nu}$, but now the role of the fluctuation metric will be played by $g^{(m+1)}_{\mu\nu}$ 
\footnote{The reason for this universality is as follows. The correction to the solution i.e., $g^{(m+1)}_{\mu\nu}$ already carries a factor of $\alpha^{m+1}$. Since we are interested in evaluating the equation at order ${\cal O}(\alpha^{m+1})$ and also if we want to collect only those terms that involves $g^{(m+1)}_{\mu\nu}$, everything else in the equation must be of zeroth order in $\alpha$. It follows that at order ${\mathcal O}(\alpha^{m+1})$ none of the $E_{\mu\nu}^{(m)}$, for $m>0$ can contribute to terms that has $g^{(m+1)}_{\mu\nu}$ and the same is true for product terms of the form $g^{(m)} \, g^{(n)}$, for $m>0$. Therefore at order ${\mathcal O}(\alpha^{m+1})$, terms that contain $g^{(m+1)}_{\mu\nu}$ can only come from the EoM at zeroth order linearized around the zeroth order solution.}. 

But the source terms (analogous to the second term in eq.\eqref{eq:2ndorder}) will not have this universal form. At order ${\cal O}(\alpha^{m+1})$, where the metric upto order ${\cal O}(\alpha^{m})$ is already fixed by solving the equations at previous orders, the source terms will be the coefficient of $\alpha^{m+1}$ in $E_{\mu\nu}$ once evaluated on $g^{(bh)}_{\mu\nu}$ as in eq.\eqref{MetExpnsn1} but corrected upto ${\mathcal O}(\alpha^m)$, that is 
$$g^{(bh)}_{\mu\nu} \, |_{\text{corrected upto} ~ {\mathcal O}(\alpha^{m})} \, = \, g^{(0)}_{\mu\nu} + \alpha ~g^{(1)}_{\mu\nu} +\cdots+\alpha^m ~g^{(m)}_{\mu\nu} \, .$$ 
Therefore, at ${\mathcal O}(\alpha^{m+1})$, the equation looks like the following 
\begin{equation} \label{EoM_at_nR}
E^{(0)}_{\mu\nu}\left[g_{\mu\nu}^{(0)}+\alpha^{m+1} \, g^{(m+1)}_{\mu\nu}\right] + E_{\mu\nu}\left[ g^{(0)}_{\mu\nu} + \alpha \, g^{(1)}_{\mu\nu} +\alpha^2 \, g^{(2)}_{\mu\nu} + \cdots + \alpha^m \, g^{(m)}_{\mu\nu}\right] = \mathcal{O}(\alpha^{m+2}) \, ,
\end{equation}  
However, all the terms contributing to the source term in eq.\eqref{EoM_at_nR} are obtained from metric functions, which are all solved until the previous order in this iterative construction. Interestingly, for our proof, we do not need the details of the source term, except for the fact that at any given order, it is a covariant tensor evaluated on a metric that solves the EoM up to the previous order 
\footnote{ We must emphasize that for this perturbative technique to work at a given order (say ${\cal O}(\alpha^{m+1})$), EoM must be solved till the previous order. This will ensure that the source term i.e., $E_{\mu\nu}$, evaluated on $(g^{(0)}_{\mu\nu} + \alpha \, g^{(1)}_{\mu\nu} +\cdots+\alpha^{m} \, g^{(m)}_{\mu\nu})$ will be non-zero only at order ${\cal O}( \alpha^{m+1})$.}.

\subsection{Zeroth law for two derivative theories of gravity, at leading order in $\alpha$-expansion}
In the previous subsection, we have described the general structure of the EoM at any given order in $\alpha$-expansion. We have seen that the starting point must be Einstein's two derivative gravity, and $g^{(0)}_{\mu\nu}$ must be an exact stationary black hole solution of the Einstein equations $E^{(0)}_{\mu\nu}$. 

It is well known that in the two derivative theory of gravity, the temperature of a stationary black hole is constant over the horizon. In this sub-section, we will review, following the strategy outlined in \S\ref{sec:Strategy}, how this can be proved in our choice of coordinate system eq.\eqref{eq:metric}. As we have already mentioned, to achieve this, we must look into the off-shell structure of the $(\tau i)$ component of the zeroth-order EoM $E^{(0)}_{\mu\nu}$. It will turn out that $E^{(0)}_{\tau i}$ is indeed of the form eq.\eqref{eq:maingoal}. When EoMs are satisfied by stationary black hole configurations, we will readily obtain eq.\eqref{eq:finalshow}. This is, therefore, enough to prove the desired result in two-derivative theories of gravity. In the following, we will argue that eq.\eqref{eq:maingoal} is indeed true. 

Let us consider two derivative theories of gravity without any matter field. To be more explicit, let us reiterate that the equation of motion eq.\eqref{eq:Eom} is
\begin{equation}
E_{\mu\nu}=E^{(0)}_{\mu\nu} \, .
\end{equation}
The metric eq.\eqref{eq:metric} upto order ${\cal O}(\alpha^0)$ in the horizon adapted coordinate system is
\begin{equation}\label{eq:metric1}
\begin{split}
ds^2&=2 d\tau~ d\rho-\rho X^{(0)}(\rho,x^i) d\tau^2+2 \rho~ \omega_i^{(0)}(\rho,x^i) d\tau d x^i+h^{(0)}_{ij}(\rho,x^i) dx^i dx^j \, .
\end{split}
\end{equation}
We would calculate $\tau i$ component of the EoM on the horizon.
\begin{equation}
\begin{split}
E_{\tau i}&=R_{\tau i}-\frac{1}{2}R g_{\tau i} \quad \Rightarrow \quad E_{\tau i}|_{\rho=0}=R_{\tau i}|_{\rho=0} \, .
\end{split}
\end{equation}
We must now compute $R_{\tau i}$ for our choice of metric eq.\eqref{eq:metric1}. Using the expression of $R_{\tau i}$ (see appendix \ref{app:Einstein}), we get
\begin{equation}
E_{\tau i}|_{\rho=0}=-\frac{1}{2}(\partial_i X^{(0)})|_{\rho=0}
\end{equation}
Using EoM, we can straightforwardly conclude 
\begin{equation}\label{eq:resultGR}
\partial_i X^{(0)}(\rho, \, x^i)|_{\rho=0}=0 \, ,
\end{equation}
which is basically eq.\eqref{eq:finalshow} upto ${\cal O}(\alpha^0)$, and therefore, implies zeroth law at the same order.


\subsection{Zeroth law for higher curvature theories of gravity at arbitrary order in $\alpha$-expansion}\label{subsec:homogeneous}
After establishing the zeroth law at the leading order in $\alpha$-expansion for two derivative theories, in this section, we aim to extend this to arbitrary higher-order in the perturbative $\alpha$ expansion. We will construct our proof using a method of induction.  It will be shown that if the temperature is constant over the horizon till order ${\mathcal O}(\alpha^n)$, then it will remain constant at order ${\mathcal O}(\alpha^{n+1})$. 
Following our strategy described in \S\ref{sec:Strategy}, and just like what we did at the zeroth order, we will again use the off-shell structure of the $(\tau i)$ component of the EoM to show this.

For convenience, let us first re-write the metric and its $\alpha$-expansion, eq.\eqref{MetExpnsn1} and eq.\eqref{eq:metexpansionrep},
\begin{equation}\label{eq:metricrep}
\begin{split}
ds^2=g^{(bh)}_{\mu\nu}dx^\mu dx^\nu
=2 \, d\tau \, d\rho-\rho \, X(\rho,x^i) d\tau^2+2 \rho \, \omega_i(\rho,x^i) d\tau d x^i+h_{ij}(\rho,x^i) dx^i dx^j \, ,
\end{split}
\end{equation}
where, 
\begin{equation}\label{eq:metexpansionrep}
\begin{split}
X(\rho,x^i)&=X^{(0)}(\rho,x^i)+\alpha \,  X^{(1)}(\rho,x^i)+\alpha^2 \,  X^{(2)}(\rho,x^i)+\cdots\\
\omega_i(\rho,x^i)&=\omega_i^{(0)}(\rho,x^i)+\alpha \, \omega_i^{(1)}(\rho,x^i)+\alpha^2 \,  \omega_i^{(2)}(\rho,x^i)+\cdots\\
h_{ij}(\rho,x^i)&=h^{(0)}_{ij}(\rho,x^i)+\alpha \,  h^{(1)}_{ij}(\rho,x^i)+\alpha^2 \,  h^{(2)}_{ij}(\rho,x^i)+\cdots \, .
\end{split}
\end{equation}

We start with the statement that we have solved the EoM accurately upto order ${\cal O}(\alpha^m)$. Also, following the same logic, we are assuming that the temperature is constant on the horizon upto order ${\cal O}(\alpha^m)$. In terms of equation it implies 
\begin{equation} \label{rel_assmptn}
\partial_i \left(X^{(0)}(\rho, \, x^i)+\alpha~ X^{(1)}(\rho, \, x^i)+\cdots+\alpha^m~ X^{(m)}(\rho, \, x^i)\right)\Big\vert_{\rho=0} =0,
\end{equation}

Given this, now, we would like to solve the EoM at order ${\mathcal O}(\alpha^{m+1})$. 
As we have discussed in the previous sub-section \S\ref{subsec:structure}, in the context of the general structure of the EoM at an arbitrary order of the $\alpha$-expansion, working at ${\mathcal O}(\alpha^{m+1})$, we will get a linear partial differential equation for the unknown $g^{(m+1)}_{\mu\nu}$. This will be a linear PDE with two types of terms; one is the homogeneous term along with another source term. In the following, we will analyze these two terms one by one. 

From eq.\eqref{EoM_at_nR} we have learned that the homogeneous part of the equation could be universally evaluated as linearisation of the Einstein tensor  $\left( \text{i.e., }E^{(0)}_{\mu\nu}\right)$ around $g^{(0)}_{\mu\nu}$, 
\begin{equation}
\text{Homogeneous part of the PDE at} \,  {\mathcal O}\left(\alpha^{m+1}\right) \, = \,  E^{(0)}_{\mu\nu}\left[g^{(0)}_{\mu\nu} + \alpha^{m+1} g^{(m+1)}_{\mu\nu}\right] + {\cal O}(\alpha^{m+2}) \, .
\end{equation}
Note in the above equation the RHS will have the leading contribution at order ${\cal O}(\alpha^{m+1})$ since by construction  $E^{(0)}_{\mu\nu}\left[g^{(0)}_{\mu\nu}\right]=0$.
For our purpose, we just need to look at the $(\tau i)$ component of the EoM. By explicit evaluation in our choice of coordinate system we could show (see appendix \ref{app:Homogeneous})
\begin{equation}\label{eq:unihomo}
\begin{split}
&E^{(0)}_{\tau i}\left[g^{(0)}_{\mu\nu} + \alpha^{m+1} g^{(m+1)}_{\mu\nu}\right]_{\rho=0} = -\left. \frac{1}{2} \, \alpha^{m+1}\, \left(\partial_{i}X^{(m+1)}\right) \right|_{\rho=0} + {\cal O}(\alpha^{m+2})
\end{split}
\end{equation}
It should be noted that in deriving eq.\eqref{eq:unihomo}, we have used the result obtained in eq.\eqref{eq:resultGR} for the leading order two derivative theory. 

Now we come to the source terms. These are the known terms at order ${\cal O}(\alpha^{m+1})$. These could be computed by evaluating the EoM, keeping terms up to order ${\cal O}(\alpha^{m+1})$, and ignoring all higher-order terms on the metric corrected up to order ${\mathcal O}(\alpha^m)$. 

Before we proceed, let us introduce a new notation here for convenience. 
For any function $Y$ that admits an $\alpha$ expansion, $Y^{(m)}$ denotes the coefficient of $\alpha^m$ and $Y^{[m]}$ denotes the expansion of $Y$ correct upto order ${\cal O}(\alpha^m)$. In other words, if $Y$ could be written as $Y = \sum_{m=0}^\infty\alpha^m \, Y^{(m)}$, then $Y^{[m]}$ denotes  \begin{equation}
Y^{[m]} \equiv \sum_{k=0}^m \alpha^k \, Y^{(k)} \, .
\end{equation}
According to this notation, then, for the corrected and truncated metric and EoM till order ${\mathcal O}(\alpha^m)$, we get
\begin{equation}
g^{[m]}_{\mu\nu} \equiv \sum_{i=0}^m \alpha^m \, g_{\mu\nu}^{(m)}, \quad \text{and} \quad E_{\mu\nu}^{[m]} \equiv \sum_{i=0}^m \alpha^m \, E_{\mu\nu}^{(m)} \, .
\end{equation}

Using this new notation, let us now write down the source term in the PDE for $g^{(m)}_{\mu\nu}$, working at  order ${\mathcal O}(\alpha^{m+1})$, as the following
\begin{equation}\label{eq:source}
\begin{split}
\text{Source terms of the PDE at} ~ {\mathcal O}(\alpha^{m+1}) \, = \left. E^{[m+1]}_{\mu\nu} \, \right\vert_{\text{evaluated on} \, g^{[m]}_{\mu\nu}} + {\mathcal O}(\alpha^{m+2})
\end{split}
\end{equation}
Note that according to our assumptions, $g^{[m]}_{\mu\nu}$ solves the EoM up to order ${\mathcal O}(\alpha^m)$. It follows that the source terms as written above will have the first non-zero contribution at order ${\mathcal O}(\alpha^{m+1})$.
As we have mentioned before, the source terms do not have any universal structure, unlike the homogeneous piece. However, for the constancy of the temperature, we need to analyze only the $(\tau i)$-component of the EoM and that too only at the horizon, i.e., $\rho =0$ hypersurface in our choice of coordinates. This will simplify our analysis.

\subsubsection{Boost symmetry for the truncated metric and vanishing of the source term}\label{subsec:boost}
In the previous sub-section, we have shown that the homogeneous part of the $(\tau i)$ component of the EoM at this order is simply proportional to $\partial_i X^{(m+1)}$, see eq.\eqref{eq:unihomo}. Therefore, what is left to be checked is that the source term in the PDE for $g^{(m+1)}_{\mu\nu}$ vanishes at this order ${\mathcal O}(\alpha^{m+1})$. In the following, we will argue that this is exactly what will turn out to be true. 

As we have described before, the source term at order ${\cal O}(\alpha^{(m+1)})$, given in eq.\eqref{eq:source}, is simply the leading piece (in terms of $\alpha$-expansion) in $E^{[m]}_{\mu\nu}$ evaluated on $g^{[m]}_{\mu\nu}$.  We should remember that, the corrected space-time metric $g^{[m]}_{\mu\nu}$ is truncated at ${\mathcal O}(\alpha^{m})$. Also, it is corrected, because, it solves the EoM till the same order. The truncated black hole metric $g_{\mu\nu}^{[m]}$, leads to the following line element
\begin{equation}\label{eq:metric1r}
\begin{split}
ds_{[m]}^2&=2 \, d\tau \, d\rho-\rho \, X^{[m]}(\rho,x^i) \, d\tau^2+2 \, \rho \, \omega_i^{[m]}(\rho,x^i) \, d\tau \, d x^i
+h^{[m]}_{ij}(\rho,x^i) \, dx^i dx^j
\end{split}
\end{equation}

As a consequence of our assumption in eq.\eqref{rel_assmptn}, the zeroth law can be assumed to be satisfied for the metric $g^{[m]}_{\mu\nu}$ till ${\mathcal O}(\alpha^{m})$. So, we are allowed to use the fact that $X^{[m]}(\rho = 0, \,x^i )$ is constant over the horizon ($\rho=0$ hypersurface), 
\begin{equation}\label{relation2}
\left. \partial_i \left( \sum_{m \le n} \left( X^{(0)}(\rho,x^i) + \alpha^m X^{(m)}(\rho,x^i) \right) \right) \right|_{\rho=0}= \, \mathcal{O}(\alpha^{m+1})  \, .
\end{equation}
This in turn enables us to ascertain that the surface gravity, $\kappa$, computed with the corrected metric till $\mathcal{O}(\alpha^{m})$, will be constant over the horizon. In other words $X^{[m]}$ could be written, by solving eq.\eqref{relation2}, as
\begin{equation}
X^{[m]}(\rho, x^i) = C^{[m]} + \rho \, F^{[m]}(\rho, x^i) \, ,
\end{equation}
where $C^{[m]}$ is a constant, and $F^{[m]}(\rho, x^i)$ is an arbitrary function of $(\rho, x^i)$, and both of them are corrected upto $\mathcal{O}(\alpha^{m})$.

Following our discussion in \S\ref{sec:BstWght} (see eq.\eqref{coordtrans1})  we would like to perform the following coordinate transformation from $x^\mu=\{\tau,\rho,x^i\}$ to $\widetilde{x}^\mu=\{v,r,x^i\}$ where, $v$ is the affine parameter along the null generator of the horizon, 
\begin{equation} \label{coordtrans2}
v=\frac{2}{C^{[m]}}\exp\left(\frac{C^{[m]}}{2}\tau\right), \quad r=\rho \, \exp\left(-\frac{C^{[m]}}{2}\tau\right) \, .
\end{equation}
The truncated metric in the new coordinates takes the form
\begin{equation}\label{eq:station1}
ds_{[m]}^2=2dv \, dr-r^2\,  F^{[m]}(C^{[m]} rv/2, x^i)dv^2+2 r \, \omega^{[m]}_i(C^{[m]}rv/2, x^i)dv \, dx^i+h^{[m]}_{ij}(C^{[m]}rv/2, x^i)dx^i dx^j
\end{equation}
Now, $E^{[m]}_{\mu\nu}$ is just a covariant tensor of rank two, constructed out of appropriate contractions of the product of Riemann tensors and/or their covariant derivatives. So, without knowing any details about it, we could tell how its component would transform under the above-mentioned coordinate transformation. By which we mean that $\widetilde{E}^{[m]}_{\mu\nu}$ in the new coordinates will be related to $E^{[m]}_{\mu\nu}$ in the old coordinates, as follows
\begin{equation}
E^{[m]}_{\mu\nu} (\rho,\tau, x^i)=\frac{\partial\widetilde{x}^\alpha}{\partial x^\mu} \, \frac{\partial\widetilde{x}^\beta}{\partial x^\nu} \,  \widetilde{E}^{[m]}_{\alpha\beta}(r,,v,x^i) \, .
\end{equation}
For our purpose we just need to study the $\tau i$-component of $E^{[m]}_{\mu\nu} (\rho,\tau, x^i)$. Also, we can readily obtain the relevant components of $\frac{\partial\widetilde{x}^\alpha}{\partial x^\mu}$ from eq.\eqref{coordtrans2}, 
\begin{equation}
{\partial v \over \partial \tau} = \exp\left(\frac{C^{[m]}}{2}\tau\right), \quad \text{and} \quad {\partial r \over \partial \tau} = -\rho\frac{C^{[m]}}{2}\exp\left(-\frac{C^{[m]}}{2}\tau\right) \, .
\end{equation}
Using them we obtain 
\begin{equation} \label{EtiEvi}
\begin{split}
\left. E^{[m]}_{\tau i}(\rho,\tau, x^i) \right|_{\rho=0} =\left. \exp\left(\frac{C^{[m]}}{2}\tau\right)\widetilde{E}^{[m]}_{vi} (r,,v,x^i)\right|_{r=0}
\end{split}
\end{equation}
It is important to note that we have obtained $E^{[m]}_{\tau i}$ is proportional to $\widetilde{E}^{[m]}_{vi}$ when evaluated on the horizon. To decide about $\widetilde{E}^{[m]}_{vi}$ in the new coordinate system $(r,,v,x^i)$ we can directly use the boost-invariance of the metric eq.\eqref{eq:metric1r}. 
As we can see that $\widetilde{E}^{[m]}_{vi}$ contains one extra lower $v$-index compared to $r$-index. According to the arguments due to this boost-symmetry the boost-weight assigned to $\widetilde{E}^{[m]}_{vi} $ comes out to be $+1$. Therefore, if we compute $\widetilde{E}^{[m]}_{vi}$ on the stationary metric eq.\eqref{eq:station} at $r=0$ it will simply vanish. This in turn shows that, by using eq.\eqref{EtiEvi}, in our old $(\rho,\tau, x^i)$ coordinates $E^{[m]}_{\tau i}$ also vanishes. 

Therefore, we have now established the fact that at $\mathcal{O}(\alpha^{m+1})$ the source term contribution to the PDE for $g^{(m+1)}_{\mu\nu}$ vanishes. The homogeneous piece is the only contribution and that too is of the form argued in eq.\eqref{eq:unihomo}. With this we have also successfully demonstrated that
\begin{equation} \label{finresult}
\partial_i X^{(k)}(\rho, \, x^i)|_{\rho=0} =0, \quad \text{for} \quad k = (m+1) \, , 
\end{equation}
once we start with the assumption of $\partial_i X^{(k)}(\rho, \, x^i)|_{\rho=0} =0$ for $k \leq m$. Finally, by method of induction, we, therefore also prove that, starting with a positive result in the leading two derivative gravity, the zeroth law is true upto all orders in the perturbative expansion in the higher derivative coupling $\alpha$.

\section{Discussions} \label{sec:disco}
In this paper, we have worked out a proof for the zeroth law of black hole thermodynamics in diffeomorphism invariant theories of gravity. Our analysis crucially depends on the fact that we consider only such theories of gravity where arbitrary higher derivative theories of gravity augment the leading two derivative theory of general relativity. We assumed that the higher derivative coupling (denoted by $\alpha$ in this paper) could be taken to zero in a smooth limit leaving us with the leading two derivative theory. This, in turn, allows us to organize our analysis in a perturbative expansion in the higher derivative coupling $\alpha$. For example, suppose we have an exact solution in the form of a black hole metric of the equations of motion coming from the two-derivative Einstein's equation. We can correct this solution in that perturbation scheme and expect to obtain the corresponding black hole solution in the higher derivative theory of gravity. The zeroth law is a statement about stationary configurations. We used a particular coordinate system, like choosing a particular gauge, to write down the space-time metric of a stationary black hole. Since the temperature of a black hole is identified with the geometric quantity called surface gravity, working within our choice of metric gauge, our main aim was to prove that the surface gravity is constant over the horizon for stationary black holes. We want to stress here that for our construction of the proof we did not need to use any extra symmetry, we have only used the boost-symmetry which follows from stationarity. 

The crucial ingredient in our construction for the proof was to use specific components of the equations of motion (EoM). We expanded the EoM order by order in a perturbation series in the higher derivative coupling $\alpha$, with the leading term (with $\alpha$ = 0) being Einstein's equation. The metric was also expanded in a similar perturbative expansion in $\alpha$, with the leading order term being the stationary black hole solution in Einstein gravity. We followed a method of induction for the proof. First, we showed that the components of EoM have the desired off-shell structure needed for the proof to go through at the leading order in Einstein's gravity. Then we assumed that this is true at the $n$-th order, and we argued that it should be satisfied at the following order in $\alpha$-expansion. 

Working at $\mathcal{O} (\alpha^{m+1})$, once we assumed that the zeroth law is satisfied at the previous order, i.e., till $\mathcal{O} (\alpha^{m})$, we made use of a specific residual gauge symmetry to perform a coordinate transformation. In this new coordinate system, the coordinate along the null generators of the horizon happens to be affinely parametrized, and the new metric enjoys a symmetry called the boost-symmetry of the stationary black holes. Using this symmetry, we could predict the structure of the components of the EoM without knowing its explicit form. In other words, we viewed the EoM at $\mathcal{O} (\alpha^{m+1})$ as a covariant tensor built out of the metric components corrected till $\mathcal{O} (\alpha^{m})$ to satisfy the zeroth law. Then, by knowing how the EoM for any arbitrary higher derivative theory should transform under the coordinate transformation (boost transformation), we could prove that it indeed has the required structure to satisfy the zeroth law at $\mathcal{O} (\alpha^{m+1})$. 

It is essential to highlight that this particular boost symmetry can be used only when the metric can be cast in the new coordinate system we mentioned above. This was crucially used in constructing the entropy current, using which the local version of the second law was argued for arbitrary diffeomorphism invariant theory of gravity. We also understood that one could write down the stationary black hole metric in these new coordinates if the zeroth law is satisfied. This was an important assumption in constructing the entropy current in \cite{Bhattacharya:2019qal} and \cite{Bhattacharyya:2021jhr}. Therefore, our proof of zeroth law in this paper justifies this important assumption that was made in those works aimed at proving the second law. 

Another critical point in constructing our proof in this paper is the assumption that it only applies to such theories where, in an appropriate $\alpha$ expansion, the leading order piece has to be Einstein's two derivative theory. It was one significant input in our proof. However, we have not argued that there cannot be any other proof that will not require this assumption of starting the perturbation series from Einstein's gravity. From the perspective of a UV complete theory of quantum gravity, it is pretty natural to expect that the low energy effective theories following from any quantum theory of gravity would organize themselves in such a perturbative framework starting with two derivative general relativity. However, it is exciting to note that without having access to the details of how UV completion is achieved and staying entirely within a low energy perspective, principles like the laws of black hole thermodynamics also hint toward general relativity as the leading theory in a perturbative framework.

Although, our proof is perturbative in higher derivative coupling constant $\alpha$, we want to stress that it works up to arbitrary order in the perturbative expansion. With this statement, we might hope that finding a proof of zeroth law for theories non-perturbatively connected to general relativity will be worth exploring. We leave that for future work.

\acknowledgments 
We want to thank Prateksh Dhivakar for collaboration at the initial stage. We would also like to thank Diptarka Das, Suchetan Das, Prateksh Dhivakar, Suman Kundu, Alok Laddha, Mangesh Mandlik, Milan Patra, Shuvayu Roy, Sudipta Sarkar for useful discussions. PB would like to acknowledge the support provided by the grants: SPO/SERB/PHY/2017504 and SB/SJF/2019-20/08. We would finally like to thank the people of India for their steady support for research in basic sciences.

\appendix

\section{Computing the Christoffel symbols and the surface gravity for the metric eq.(\ref{eq:metric})}\label{app:kappa}
\subsection{Computing the Christoffel symbols} \label{app:chris}
In this appendix, we will calculate the Christoffel symbols  for the metric eq.\eqref{eq:metric} upto order ${\cal O}(\alpha)$.
\begin{equation}
\begin{split}
ds^2&=2 d\tau~ d\rho-\rho \left[X^{(0)}(\rho,x^i)+\alpha X^{(1)}(\rho,x^i)\right] d\tau^2+2 \rho \left[\omega_i^{(0)}(\rho,x^i)+\alpha\omega_i^{(1)}(\rho,x^i)\right] d\tau d x^i\\
&+\left[h^{(0)}_{ij}(\rho,x^i)+\alpha h^{(1)}_{ij}(\rho,x^i)\right] dx^i dx^j
\end{split}
\end{equation}
Different components of the metric are
\begin{equation}
\begin{split}
&g_{\tau\tau}=-\rho \left[X^{(0)}(\rho,x^i)+\alpha X^{(1)}(\rho,x^i)\right],~~g_{\tau\rho}=1,~~g_{\tau i}=\rho \left[\omega_i^{(0)}(\rho,x^i)+\alpha\omega_i^{(1)}(\rho,x^i)\right],\\
&g_{\rho\rho}=0,~~g_{\rho i}=0,~~g_{ij}=\left[h^{(0)}_{ij}(\rho,x^i)+\alpha h^{(1)}_{ij}(\rho,x^i)\right]
\end{split}
\end{equation}
Different components of the inverse metric up to order ${\cal O}(\alpha)$ are
\begin{equation} \label{eq:metinvcomp}
\begin{split}
&g^{\tau\tau}=0,~~g^{ \tau\rho}=1,~~g^{\tau i}=0,\\
&g^{\rho\rho}=\rho\left[X^{(0)}+\alpha X^{(1)}\right]+\rho^{2}h_{(0)}^{ij}\omega^{(0)}_i\omega^{(0)}_j+\alpha\rho^{2}\left[2h^{ij}_{(0)}\omega^{(0)}_i \omega^{(1)}_j-h_{(1)}^{ij}\omega^{(0)}_{i}\omega^{(0)}_{j}\right] \\
&g^{\rho i}=-\rho\left[h^{ij}_{(0)}\omega_{j}^{(0)}+\alpha\left(h_{(0)}^{ij}\omega^{(1)}_{j}-h_{(1)}^{ij}\omega^{(0)}_{j}\right)\right],~~ g^{ij}=h^{ij}_{(0)} - \alpha h^{ij}_{(1)}
\end{split}
\end{equation}
where, $h_{(0)}^{ij}$ is defined as $h_{(0)}^{ik}h^{(0)}_{kj}=\delta^i_j$ and  $h_{(1)}^{ij}$ is defined as $h_{(1)}^{ij}=h_{(0)}^{im}h_{(0)}^{jn}h^{(1)}_{mn}$.\\
Now we will compute different components of Christoffel symbols. We would require the expression of one component of the Christoffel symbol $\Gamma^\rho_{i\tau}$ off the horizon. The expression of $\Gamma^\rho_{i\tau}$ up to order ${\cal O}(\rho)$ is
\begin{equation}
\Gamma^\rho_{i\tau}=-\frac{\rho}{2}\partial_i \left(X^{(0)}+\alpha X^{(1)}\right)-\frac{1}{2}\rho\left(X^{(0)}\omega_i^{(0)}+\alpha X^{(1)} \omega^{(0)}_{i}+ \alpha X^{(0)} \omega^{(1)}_{i}\right)
\end{equation}
The rest of the components are on the horizon
\begin{equation}\label{eq:Gamma}
\begin{split}
&\Gamma^\rho_{\rho\tau}=-\frac{1}{2}\left[X^{(0)}+\alpha X^{(1)}\right],~~\Gamma^\tau_{i\tau}=-\frac{1}{2}\left[\omega_i^{(0)}+\alpha \omega^{(1)}_i\right],~~\Gamma^\rho_{\rho j}=\frac{1}{2}\left[\omega_i^{(0)}+\alpha ~\omega^{(1)}_i\right],\\
&\Gamma^j_{i\tau}=0,~~\Gamma^j_{\rho\tau}=\frac{1}{2}\left[\omega^{j}_{(0)}+\alpha~ \omega_{(1)}^{j}-\alpha h_{(1)}^{jk} \omega^{(0)}_{k} \right],~~\Gamma^\rho_{ij}=0,~~\Gamma^\tau_{\tau\tau}=\frac{1}{2}\left[X^{(0)}+\alpha X^{(1)}\right],\\
&\Gamma^\tau_{\tau\rho}=0,~~\Gamma^\tau_{i\rho}=0,~~ \Gamma^j_{\tau\tau}=0,~~\Gamma^\tau_{ij}=-\frac{1}{2}\left(\partial_\rho h^{(0)}_{ij}+\alpha\partial_{\rho}h^{(1)}_{ij}\right),~~\Gamma^\rho_{\tau\tau}=0,~~\Gamma^i_{\tau\tau}=0,
\end{split}
\end{equation}
Where, $\omega^{i}_{(0)}$ and $\omega^{i}_{(1)}$ are defined as $\omega^{i}_{(0)}=h^{ij}_{(0)}\omega^{(0)}_j$ and $\omega^{i}_{(1)}=h^{ij}_{(0)}\omega^{(1)}_j$ 
\subsection{Computing the surface gravity}
The metric of the space-time is given in eq.\eqref{eq:metric} and we write it here again for convenience
\begin{equation}\label{eqmet1}
\begin{split}
ds^2=2 d\tau \, d\rho-\rho X(\rho,x^i) d\tau^2+2 \rho~ \omega_i(\rho,x^i) d\tau d x^i+h_{ij}(\rho,x^i) dx^i dx^j
\end{split}
\end{equation}
This metric admits a Killing vector $\xi = \partial_{\tau}$ with the horizon being chosen to be at $\rho = 0$. The definition of surface gravity is given by 
\begin{equation}
\kappa=\left.\sqrt{-\frac{1}{2}(\nabla_\mu \xi_\nu)(\nabla^\mu \xi^\nu)} \, \right\vert_{\rho=0} \, .
\end{equation}

We use the inverse metric expressions written in eq.\eqref{eq:metinvcomp} to obtain the following components of $\xi_\mu$,
\begin{equation}
\xi_\rho = \, 1, \, \xi_\tau = \, -\rho \, X(\rho,x^i), \, \xi_i =  \, \rho \, \omega_i (\rho,x^i)\, .
\end{equation}

Next we compute the components of $\nabla_\mu \xi_\nu$ evaluated on $\rho=0$ and the non-vanishing components are as follows
\begin{equation}
\begin{split}
\nabla_\tau \xi_\rho |_{\rho=0} &=- \, \nabla_\rho \xi_\tau |_{\rho=0}= \, - {1 \over 2}  \, X(\rho = 0,x^i) \, , \\
\nabla_\rho \xi_i |_{\rho=0} &=- \, \nabla_i \xi_\rho |_{\rho=0}= \, {1 \over 2} \,  \omega_i(\rho = 0,x^i) \, ,
\end{split}
\end{equation}

Using these, we obtain
\begin{equation}
(\nabla_\mu \xi_\nu)(\nabla^\mu \xi^\nu) \, |_{\rho=0} \, = 2 g^{\tau\rho}g^{\tau\rho} \, (\nabla_\tau \xi_\rho)\, (\nabla_\rho \xi_\tau)\, |_{\rho=0} =- {1 \over 2}  \, X^2(\rho = 0,x^i)
\end{equation}

Finally, we obtain the surface gravity as the following 
\begin{equation}
\kappa = \left. {1\over 2} X(\rho,x^i) \, \right \vert_{\rho=0}  \, .
\end{equation}

\section{Few details regarding the boost weight of covariant tensors} \label{app:ReviewBoostWeight}
In this appendix we aim to provide some more detail regarding the boost invariance of the stationary metric written in eq.\eqref{eq:metricAFF}. We write the metric here again for convenience, 
\begin{equation} \label{eq:metricAFFr}
ds^2=\tilde{g}^{(bh)}_{\mu\nu} \, dx^\mu dx^\nu
=2  \, dv \, dr-r^2 \,  X(rv, \, x^i) \, dv^2+2 \, r \,  \omega_i(rv, \, x^i)  \, dv  \, d x^i+h_{ij}(rv, \, x^i)  \, dx^i  \, dx^j \, .
\end{equation}
The vector $\xi$ defined in eq.\eqref{killingvector} ,generates Killing symmetry of the stationary background with the metric in eq.\eqref{eq:metricAFFr}. Due to this, as we have already mentioned before, when we operate Lie derivative with respect to $\xi$ on any covariant tensor constructed out of the stationary metric eq.\eqref{eq:metricAFFr}, will vanish. To be more precise, acting with the Lie derivative with respect to $\xi$, on a covariant tensor, say $\mathcal{B}_{\mu_1 \mu_2 \cdots \mu_k}$ with all lowered indices, will produce the following,
\begin{equation}\label{Lxi}
\begin{split}
{\cal L}_\xi \mathcal{B}_{\mu_1 \mu_2 \cdots \mu_k}= \, &\xi^\beta\partial_\beta \mathcal{B}_{\mu_1 \mu_2 \cdots \mu_k}+ \left(\partial_{\mu_1}\xi^{\beta}\right)\mathcal{B}_{\beta \mu_2 \cdots \mu_k}+ \left(\partial_{\mu_2}\xi^{\beta}\right)\mathcal{B}_{\mu_1 \beta \cdots \mu_k}+\cdots\\
& + \left(\partial_{\mu_k}\xi^{\beta}\right)\mathcal{B}_{\mu_1 \mu_2 \cdots \beta}\, .
\end{split}
\end{equation}
Furthermore, when we evaluate this for the metric eq.\eqref{eq:metricAFFr}, and with $xi$ given in eq.\eqref{killingvector}, we will get
\begin{equation}\label{Lxi1}
\begin{split}
{\cal L}_\xi \mathcal{B}_{\mu_1 \mu_2 \cdots \mu_k}=\,  \left[ w +(v\partial_v -r\partial_r) \right]\mathcal{B}_{\mu_1 \mu_2 \cdots \mu_k}\, ,
\end{split}
\end{equation}
where $w$ is the boost weight of $\mathcal{B}_{\mu_1 \mu_2 \cdots \mu_k}$ and from eq.\eqref{Lxi1} we can also confirm that $w$ counts the excess number of lower $v$ indices compared to lower $r$ indices in $\mathcal{B}_{\mu_1 \mu_2 \cdots \mu_k}$. Following this argument, it is also obvious that the $vi$-component of EoM, $E_{vi}$ will have boost weight equal to $+1$, and hence, will vanish for stationary configurations when evaluated on the horizon. This is the main ingredient that we have used in \S\ref{subsec:boost}. 

Before we conclude this appendix let us summarise the useful points that we should remember while using the boost weight analysis, 
\begin{enumerate}
\item We should think about any component of a covariant tensor to have a structure with some number of $\partial_r$, $\partial_v$ and $\nabla_i$ operators acting on the metric coefficients in eq.\eqref{eq:metricAFFr}: ($X$, $\omega_i$, and $h_{ij}$) or product of such structures.
\item The boost weight of any covariant tensor can be obtained by looking at the factor $w$ in eq.\eqref{Lxi1}, when a Lie derivative ${\cal L}_\xi$, with respect to $\xi \left(= v\partial_v - r\partial_r\right)$, acts on it.
\item Any expression with positive boost weight will vanish when evaluated on the horizon for a stationary metric.
\end{enumerate}
For more details we refer the reader to section-(2.3) and Appendix-B of \cite{Bhattacharyya:2021jhr}.

\section{More detailed calculation for Einstein's gravity}\label{app:Einstein}
In this appendix, we will calculate $\tau i$ component of equation of motion $E_{\tau i}^{(0)}$ for Einstein's gravity.
\begin{equation}
R_{\tau i}={R^\tau}_{\tau\tau i}+{R^\rho}_{\tau\rho i}+{R^j}_{\tau j i}
\end{equation}
Using the expressions of Christoffel symbols computed in Appendix-(\ref{app:chris}), we can calculate different components of Riemann tensor upto order ${\cal O}(\alpha^0)$
\begin{equation}
\begin{split}
{R^\tau}_{\tau\tau i}|_{\rho=0}&=\partial_\tau \Gamma^\tau_{i\tau}-\partial_i\Gamma^\tau_{\tau\tau}+\Gamma^\tau_{\tau E}\Gamma^E_{i\tau}-\Gamma^\tau_{iE}\Gamma^E_{\tau\tau}\\
&=-\frac{1}{2}\partial_i X^{(0)}
\end{split}
\end{equation}
\begin{equation}
\begin{split}
{R^\rho}_{\tau\rho i}|_{\rho=0}&=\partial_\rho \Gamma^\rho_{i\tau}-\partial_i\Gamma^\rho_{\rho\tau}+\Gamma^\rho_{\rho E}\Gamma^E_{i\tau}-\Gamma^\rho_{iE}\Gamma^E_{\rho\tau}\\
&=0
\end{split}
\end{equation}
\begin{equation}
\begin{split}
{R^j}_{\tau j i}|_{\rho=0}&=\partial_j \Gamma^j_{i\tau}-\partial_i\Gamma^j_{j\tau}+\Gamma^j_{jE}\Gamma^E_{i\tau}-\Gamma^j_{iE}\Gamma^E_{j\tau}\\
&=0
\end{split}
\end{equation}
Finally we get
\begin{equation}
E_{\tau i}|_{\rho=0}=R_{\tau i}|_{\rho=0}=-\frac{1}{2}(\partial_i X^{(0)})|_{\rho=0}
\end{equation}

\section{Calculation of the homogeneous part}\label{app:Homogeneous}
In this appendix, we will derive the expression of the homogeneous part eq.\eqref{eq:unihomo}. As has been discussed in sub-section \S\ref{subsec:structure}, we have to linearize $E^{(0)}_{\mu\nu}$ around $g^{(0)}_{\alpha\beta}$.
We have to calculate $E^{(0)}_{\mu\nu}\left[g^{(0)}_{\alpha\beta}+\delta g_{\alpha\beta}\right]$, where, we will treat $\delta g_{\alpha\beta}\equiv \alpha^{m+1}g_{\alpha\beta}^{(m+1)}$ as linearized perturbations around $g^{(0)}_{\alpha\beta}$. $E^{(0)}_{\mu\nu}$ is the Einstein's tensor
\begin{equation}
E^{(0)}_{\mu\nu}=R_{\mu\nu}-\frac{1}{2}R g_{\mu\nu}
\end{equation}
As, $g^{(0)}_{\mu\nu}$ is an exact solution of $E^{(0)}_{\mu\nu}$
\begin{equation}
E^{(0)}_{\mu\nu}\left[g^{(0)}_{\alpha\beta}+\delta g_{\alpha\beta}\right]\equiv \delta E^{(0)}_{\mu\nu}=\delta R_{\mu\nu}-\frac{1}{2}g^{(0)}_{\mu\nu}\delta R -\frac{1}{2}R^{(0)}\delta g_{\mu\nu}
\end{equation}
$R^{(0)}$ is the Ricci scalar evaluated on the metric $g^{(0)}_{\mu\nu}$. We have to calculate $\tau i$ component of the above equation at $\rho=0$. We can compute $\delta E^{(0)}_{\tau i}$ off the horizon, but for our purpose that is not required.
\begin{equation}
\delta E^{(0)}_{\tau i}|_{\rho=0}=\delta R_{\tau i}-\frac{1}{2}g^{(0)}_{\tau i}\delta R -\frac{1}{2}R^{(0)}\alpha^{m+1}g^{(m+1)}_{\tau i}=\delta R_{\tau i}
\end{equation}
Since we have denoted the coordinates by $\{\tau,\rho,x^i\}$, for notational convenience, instead of using $\mu,\nu$ we will be denoting the spacetime coordinates by $\{A,B,C...\}$. We will be using this notation only for this appendix. If we calculate the Christoffel symbols on $g_{AB}^{(0)}+\alpha^{m+1}g^{(m+1)}_{AB}$ we can decompose it as follows
\begin{equation}
\Gamma^A_{BC}=\bar{\Gamma}^A_{BC}+\delta\Gamma^A_{BC}
\end{equation}
where $\bar{\Gamma}^A_{BC}$ is the Christoffel symbols for $g^{(0)}_{\mu\nu}$. Linearized Ricci tensor is
\begin{equation}
\begin{split}
\delta R_{AB}&=\nabla_D\delta \Gamma^D_{AB}-\nabla_B\delta \Gamma^D_{AD}\\
\end{split}
\end{equation}
We can very easily read-off the expressions of $\bar{\Gamma}^A_{BC}$ and $\delta \Gamma^A_{BC}$ from eq.\eqref{eq:Gamma}.
\begin{equation}
\delta R_{AB}=\partial_D\left(\delta \Gamma^D_{AB}\right)+\bar{\Gamma}^D_{DE}\delta\Gamma^E_{AB}-\bar{\Gamma}^E_{DA}\delta \Gamma^D_{EB}-\bar{\Gamma}^E_{DB}\delta \Gamma^D_{AE}-\partial_B \delta \Gamma^D_{AD}+\bar{\Gamma}^E_{BA}\delta\Gamma^D_{ED}
\end{equation}
\begin{equation}\label{eq:deltaRti}
\delta R_{\tau i}=\partial_D\left(\delta \Gamma^D_{\tau i}\right)+\bar{\Gamma}^D_{DE}\delta\Gamma^E_{\tau i}-\bar{\Gamma}^E_{D\tau}\delta \Gamma^D_{Ei}-\bar{\Gamma}^E_{Di}\delta \Gamma^D_{\tau E}-\partial_i \delta \Gamma^D_{\tau D}+\bar{\Gamma}^E_{i \tau}\delta\Gamma^D_{ED}
\end{equation}
Now, we will compute different terms of the above equation on $\rho=0$ separately
\begin{equation}\label{eq:term1}
\begin{split}
\partial_D \delta \Gamma^D_{\tau i}&=\partial_\rho \delta \Gamma^\rho_{\tau i}+\partial_\tau \delta \Gamma^\tau_{\tau i}+\partial_j \delta \Gamma^j_{\tau i}\\
&=-\frac{1}{2}\alpha^{m+1}\partial_i\left(X^{(m+1)}\right)-\frac{1}{2}\alpha^{m+1}\left(X^{(m+1)}\omega_i^{(0)}+X^{(0)}\omega_i^{(m+1)}\right)
\end{split}
\end{equation}
\begin{equation}
\begin{split}
\bar{\Gamma}^D_{DE}\delta\Gamma^E_{\tau i}&=\bar{\Gamma}^D_{D\tau}\delta\Gamma^\tau_{\tau i}\\
&=\left(\bar{\Gamma}^\tau_{\tau\tau}+\bar{\Gamma}^\rho_{\rho\tau}\right)\delta\Gamma^\tau_{\tau i}\\
&=\left(\frac{1}{2}X^{(0)}-\frac{1}{2}X^{(0)}\right)\left(-\frac{1}{2}\alpha^{m+1}\omega_i^{(m+1)}\right)\\
&=0
\end{split}
\end{equation}
\begin{equation}
\begin{split}
\bar{\Gamma}^E_{D\tau}\delta \Gamma^D_{Ei}&=\bar{\Gamma}^\tau_{D\tau}\delta\Gamma^D_{\tau i}+\bar{\Gamma}^\rho_{D\tau}\delta \Gamma^D_{\rho i}+\bar{\Gamma}^j_{D\tau}\delta \Gamma^D_{ji}\\
&=\bar{\Gamma}^\tau_{\tau\tau}\delta\Gamma^\tau_{\tau i}+\bar{\Gamma}^\rho_{\rho\tau}\delta \Gamma^\rho_{\rho i}+\bar{\Gamma}^j_{\rho\tau}\delta \Gamma^\rho_{ji}\\
&=\frac{1}{2}X^{(0)}\left(-\frac{1}{2}\alpha^{m+1}\omega_i^{(m+1)}\right)-\frac{1}{2}X^{(0)}\frac{1}{2}\alpha^{m+1}\omega_i^{(m+1)}\\
&=-\frac{1}{2}X^{(0)}\alpha^{m+1}\omega_i^{(m+1)}
\end{split}
\end{equation}
\begin{equation}
\begin{split}
\bar{\Gamma}^E_{Di}\delta\Gamma^D_{\tau E}&=\bar{\Gamma}^\tau_{Di}\delta\Gamma^D_{\tau \tau}+\bar{\Gamma}^\rho_{Di}\delta\Gamma^D_{\tau \rho}+\bar{\Gamma}^j_{Di}\delta\Gamma^D_{\tau j}\\
&=\bar{\Gamma}^\tau_{\tau i}\delta\Gamma^\tau_{\tau \tau}+\left(\bar{\Gamma}^\rho_{\rho i}\delta\Gamma^\rho_{\tau \rho}+\bar{\Gamma}^\rho_{ji}\delta\Gamma^j_{\tau \rho}\right)+\bar{\Gamma}^j_{\tau i}\delta\Gamma^\tau_{\tau j}\\
&=-\frac{1}{2}\omega_i^{(0)}\frac{1}{2}\alpha^{m+1}X^{(m+1)}+\frac{1}{2}\omega_i^{(0)}\left(-\frac{1}{2}\alpha^{m+1}X^{(m+1)}\right)\\
&=-\frac{1}{2}\omega_i^{(0)}\alpha^{m+1}X^{(m+1)}
\end{split}
\end{equation}
\begin{equation}
\begin{split}
\partial_i \delta \Gamma^D_{\tau D}&=\partial_i\left(\delta \Gamma^\rho_{\tau \rho}+\delta \Gamma^\tau_{\tau \tau}\right)\\
&=\partial_i\left(-\frac{1}{2}\alpha^{m+1}X^{(m+1)}+\frac{1}{2}\alpha^{m+1}X^{(m+1)}\right)\\
&=0
\end{split}
\end{equation}
\begin{equation}\label{eq:term8}
\begin{split}
\bar{\Gamma}^E_{i \tau}\delta\Gamma^D_{ED}&=\bar{\Gamma}^\tau_{i \tau}\delta\Gamma^D_{\tau D}\\
&=\bar{\Gamma}^\tau_{i \tau}\left(\delta\Gamma^\rho_{\tau \rho}+\delta\Gamma^\tau_{\tau \tau}\right)\\
&=-\frac{1}{2}\omega_i^{(0)}\left(-\frac{1}{2}\alpha^{m+1}X^{(m+1)}+\frac{1}{2}\alpha^{m+1}X^{(m+1)}\right)\\
&=0
\end{split}
\end{equation}
Substituting eq.\eqref{eq:term1} - eq.\eqref{eq:term8} in eq.\eqref{eq:deltaRti} we get
\begin{equation}
\delta R_{\tau i}=-\frac{1}{2}\alpha^{m+1}\partial_i\left(X^{(m+1)}\right)
\end{equation}

\bibliographystyle{JHEP}
\bibliography{ZerothLaw}

\end{document}